\documentclass[11pt,twoside,a4paper,cmspaper,final,collab]{cms-tdr}
\def\svnVersion{ba5c897}\def\svnDate{2026/07/21}\def\cmsCernNoTag{CERN-EP-2025-176}\def\cmsCernDate{\today}\def\cmsMessage{Published in Physical Review D as \href{http://dx.doi.org/10.1103/36cv-vst3}{\doi{10.1103/36cv-vst3}.}}
\begin{document}\cmsNoteHeader{BPH-21-007}

\newcommand{\fu}        {\ensuremath{f_\PQu}\xspace}
\newcommand{\fd}        {\ensuremath{f_\PQd}\xspace}
\newcommand{\fs}        {\ensuremath{f_\PQs}\xspace}
\newcommand{\fsfu}      {\ensuremath{\fs/\fu}\xspace}
\newcommand{\fsfd}      {\ensuremath{\fs/\fd}\xspace}
\newcommand{\fdfu}      {\ensuremath{\fd/\fu}\xspace}
\newcommand\mystrut{\rule{0pt}{0.9\normalbaselineskip}}

\newcommand{\PKstn}{\HepParticle{\PK}{}{\ast 0}\xspace}
\newcommand{\PKstp}{\HepParticle{\PK}{}{\ast +}\xspace}
\newcommand\Pphi{\ensuremath{\PGf}\xspace}

\newcommand\Bph{\ensuremath{\PBp\!\to\!\PGpp\PADz(\PKp\PGpm)}\xspace}
\newcommand\Bzh{\ensuremath{\PBz\!\to\!\PGpp\PDm(\PKp\PGpm\PGpm)}\xspace}
\newcommand\Bsh{\ensuremath{\PBzs\!\to\!\PGpp\PDms(\PGpm\Pphi(\PKp\PKm))}\xspace}

\newcommand\Bphs{\ensuremath{\PBp\!\to\!\PGpp\PADz}\xspace}
\newcommand\Bzhs{\ensuremath{\PBz\!\to\!\PGpp\PDm}\xspace}
\newcommand\Bshs{\ensuremath{\PBzs\!\to\!\PGpp\PDms}\xspace}
\newcommand\BzhKs{\ensuremath{\PBz\!\to\!\PKp\PDm}\xspace}

\newcommand\BphK{\ensuremath{\PBp\!\to\!\PKp\PADz(\PKp\PGpm)}\xspace}
\newcommand\BzhK{\ensuremath{\PBz\!\to\!\PKp\PDm(\PKp\PGpm\PGpm)}\xspace}
\newcommand\BshK{\ensuremath{\PBzs\!\to\!\PKp\PDms(\PGpm\Pphi(\PKp\PKm))}\xspace}

\newcommand\Bpc{\ensuremath{\PBp\!\to\!\JPsi(\MM)\PKp}\xspace}
\newcommand\Bzc{\ensuremath{\PBz\!\to\!\JPsi(\MM)\PKstn(\PKp\PGpm)}\xspace}
\newcommand\Bsc{\ensuremath{\PBzs\!\to\!\JPsi(\MM)\Pphi(\PKp\PKm)}\xspace}

\newcommand\Bpcs{\ensuremath{\PBp\!\to\!\JPsi\PKp}\xspace}
\newcommand\Bzcs{\ensuremath{\PBz\!\to\!\JPsi\PKstn}\xspace}
\newcommand\Bscs{\ensuremath{\PBzs\!\to\!\JPsi\Pphi}\xspace}

\newcommand\Dzh{\ensuremath{\PADz\!\to\!\PKp\PGpm}\xspace}
\newcommand\Dmh{\ensuremath{\PDm\!\to\!\PKp\PGpm\PGpm}\xspace}
\newcommand\Dsh{\ensuremath{\PDms\!\to\!\PGpm\Pphi}\xspace}
\newcommand\phiKK{\ensuremath{\Pphi\!\to\!\PKp\PKm}\xspace}
\newcommand\KsKpi{\ensuremath{\PKstn\!\to\!\PKp\PGpm}\xspace}
\newcommand\DzKpi{\ensuremath{\PDz\!\to\!\PKm\PGpp}\xspace}
\newcommand\DmKpipi{\ensuremath{\PDm\!\to\!\PKp\PGpm\PGpm}\xspace}

\newcommand\Vud{\ensuremath{V_{\PQu\PQd}}\xspace}
\newcommand\Vus{\ensuremath{V_{\PQu\PQs}}\xspace}
\newcommand\fK{\ensuremath{f_{\PK}}\xspace}
\newcommand\fpi{\ensuremath{f_{\PGp}}\xspace}
\newcommand\Na{\ensuremath{N_{\mathrm{a}}}\xspace}
\newcommand\NF{\ensuremath{N_{\mathrm{F}}}\xspace}
\newcommand\tBz{\ensuremath{\tau_{\!\PBz}}\xspace}
\newcommand\tBzs{\ensuremath{\tau_{\!\PBzs}}\xspace}
\newcommand\Ncorr{\ensuremath{N_{\text{corr}}}\xspace}

\newcommand\lumunit{\ensuremath{[10^{34}\unit{cm}^{-2}\unit{s}^{-1}]}\xspace}
\newcommand\mutrig{\ensuremath{\PGm_{\text{trig}}}\xspace}
\newcommand\ptm{\ensuremath{p^\PGm_{\mathrm{T}}}\xspace}
\newcommand\ipsigtrg{\ensuremath{\text{IP}^{\text{beam}}_{\text{sig}}}\xspace}
\newcommand\ipsig{\ensuremath{\text{IP}_{\text{sig}}}\xspace}

\newcommand\BR{\ensuremath{\mathcal{B}}\xspace}
\newcommand\R{\ensuremath{\mathcal{R}}\xspace}
\newcommand\Rs{\ensuremath{\mathcal{R}_\PQs}\xspace}
\newcommand\Rsd{\ensuremath{\mathcal{R}_\PQs^\PQd}\xspace}
\newcommand\absy{\ensuremath{\abs{y}}\xspace}

\newlength{\figwid}
\ifthenelse{\boolean{cms@external}}
{
  \providecommand{\cmsLeft}{upper\xspace}
  \providecommand{\cmsRight}{lower\xspace}
  \setlength{\figwid}{0.48\textwidth}
  }{
    \providecommand{\cmsLeft}{left\xspace}
    \providecommand{\cmsRight}{right\xspace}
    \setlength{\figwid}{0.66\textwidth}
}

\cmsNoteHeader{BPH-21-007}
\title{Measurement of \texorpdfstring{\PB}{B} meson production fraction ratios in proton-proton collisions at \texorpdfstring{$\sqrt{s} = 13\TeV$}{sqrt(s) = 13 TeV} using open-charm and charmonium decays}

\date{\today}

\abstract{Production fraction ratios of \PBp, \PBz, and \PBzs mesons are measured in proton-proton collisions at $\sqrt{s} = 13\TeV$ using a special data set recorded in 2018 with high-rate triggers designed to collect an unbiased sample of $10^{10}$ \PQb hadrons with the CMS experiment at the LHC\@. These data allow the study of the open-charm decays of \PB mesons ($\PB_{(\PQs)} \!\to\! \PGp\PD_{(\PQs)}$) where the $\PD$ meson decays into fully hadronic final states. By utilizing known branching fractions and precise theoretical calculations, production fraction ratios as functions of \PB meson transverse momentum (\pt) and rapidity ($y$) are measured using open-charm decays in the kinematic range of $8 < \pt < 60\GeV$ and $\absy < 2.25$. In addition, the same data set is used to measure the relative production fraction ratios with the charmonium decay channels ($\PB_{(\PQs)} \!\to\! \PX\JPsi$ with \PX indicating a $\PKp$, $\PKstP{892}{}^0$, or $\PGfP{1020}$ meson) where the $\JPsi$ meson decays into a pair of muons.  The open-charm results are used to normalize the relative production fraction ratios obtained from the charmonium samples.  Measurements of the ratios of branching fractions of \PB meson decays to charmonium and open-charm final states are also reported, which will improve the world-average values of these ratios.  Finally, we test isospin invariance in \PB meson production in proton-proton collisions and observe that it holds within the experimental precision.}

\hypersetup{%
pdfauthor={CMS Collaboration},%
pdftitle={Measurement of B meson production fraction ratios in proton-proton collisions at sqrt(s) = 13 TeV using open-charm and charmonium decays},%
pdfsubject={CMS},%
pdfkeywords={CMS, BPH, fragmentation fractions, data parking}} 

\maketitle

\section{Introduction}

\label{sec:intro}
Studies of production fractions of various species of \PQb hadrons have been a topic of intense experimental and theoretical interest since the CERN Large Electron-Positron (LEP) collider era. These fractions, defined as the ratios of the number of produced \PQb hadron species to the total number of produced \PQb hadrons, are closely related to the \PQb quark fragmentation fractions. Specifically, the fractions \fu, \fd, and \fs are the production fractions for \PBp, \PBz, and \PBzs mesons, respectively. Here and in what follows, charge-conjugate final states are implicitly included unless otherwise indicated, and $\PK^{\ast 0}$ and $\PGf$ represent the $\PKstP{892}{}^0$ and $\PGfP{1020}$ mesons, respectively.  Production fraction ratios (PFRs), namely \fsfu, \fsfd, and \fdfu, are also of interest.  Measurements of the production fractions and PFRs have been performed in electron-positron $(\EE)$, proton-antiproton $(\Pp\Pap)$, and proton-proton $(\Pp\Pp)$ collisions at various energies. Of particular interest are measurements of the last decade conducted at the CERN LHC by the LHCb~\cite{LHCb:2011leg,LHCb:2013vfg,LHCb:2019fns,LHCb:2019lsv,LHCb:2021qbv}, ATLAS~\cite{ATLAS:2015esn}, and CMS~\cite{CMS:2022wkk} experiments, as the results are used in a variety of ways.  The \PQb hadron production fraction measurements directly enter the precision measurements of branching fractions of \PBzs meson decays, as the yields of \PBp or \PBz meson decays are used to normalize the \PBzs measurements.  The only measurements of branching fractions of \PBzs decays that are not directly dependent on the \PQb hadron production ratios have been performed by Belle on the $\PGUP{5S}$ peak and are presently limited to about 20\% precision~\cite{Belle:2012tsw}. Furthermore, the dominant systematic uncertainty in the most precise single measurement of the rare $\PBzs \!\to\! \MM$ decay, performed by the CMS Collaboration~\cite{CMS:2022mgd}, is from the uncertainty in \fsfu.

The LHCb Collaboration has studied \fsfu and \fsfd as functions of the absolute rapidity ($\absy$) and transverse momentum ($\pt$) of the \PB mesons as well as the $\Pp\Pp$ collision energy.  While no evidence has been found for a dependence on $\absy$, there is a clear dependence on both the \PB \pt and the $\Pp\Pp$ collision energy~\cite{LHCb:2011leg,LHCb:2013vfg,LHCb:2019fns,LHCb:2019lsv,LHCb:2021qbv}. The latest LHCb result~\cite{LHCb:2021qbv} finds that \fsfd decreases linearly versus \PB \pt in the $\pt < 40\GeV$ range accessible by this measurement for $\Pp\Pp$ collisions at $\sqrt{s}=13\TeV$.  A recent CMS measurement~\cite{CMS:2022wkk} using data collected at $\sqrt{s} = 13\TeV$ finds a similar dependence for \fsfu for $\pt \lesssim 18\GeV$.  By extending the measurement up to $\pt<70\GeV$, it is found that \fsfu reaches an asymptotic value around 18\GeV, remaining constant above this value.  This measurement was performed using the charmonium decay channels (\Bpc, \Bzc, and \Bsc). The present world-average value of $\BR(\PBzs \!\to\! \JPsi \Pphi)$ is dominated by the LHCb measurement~\cite{LHCb:2021qbv}, which uses the PFR measured by LHCb as an input.  As a result, while the \fsfu dependence on \PB \pt can be determined, the normalization is unknown, and therefore the CMS result is expressed in terms of the relative PFR, \Rs, defined as the ratio of efficiency-corrected yields of \PBzs to \PBp mesons, which is proportional to \fsfu.

Another area of interest is testing the limits of isospin symmetry, which postulates the same production rate for \PBp and \PBz, \ie, $\fdfu = 1$.  In most applications of PFRs, \eg, when measuring various branching fractions, the isospin symmetry in production is assumed.  While the isospin symmetry may be a good assumption in the case of isospin-singlet $\EE$ collisions at the \PZ or $\PGUP{4S}$, it is less obvious for hadronic collisions, where the production process is much more complicated. Even in $\EE$ collisions, the $\fu = \fd$ assumption may not hold at the $\PGUP{4S}$ resonance, where the majority of data are taken by \PQB factories, due to the Coulomb interaction in the $\PBp\PBm$ system near the production threshold, which is absent for the $\PBz\PABz$ system.  In fact, there are hints at about two standard deviations that $\fu \neq \fd$ for the $\PGUP{4S}$ production of \PB mesons~\cite{HFLAV:2024,Bernlochner:2023bad}.  The CMS Collaboration provided the first test of isospin invariance in $\Pp\Pp$ collisions by measuring the \fdfu ratio and demonstrating that it is flat in the entire $12 < \pt < 70\GeV$ range of the measurement and consistent with unity $(\fdfu = 0.998 \pm 0.063)$ within the 6\% precision of the measurement~\cite{CMS:2022wkk}.

\begin{figure}[bhtp]
\centering
\includegraphics[width=0.48\textwidth]{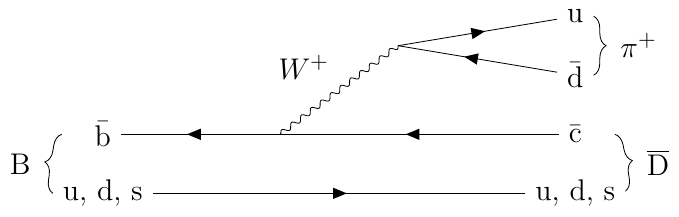}
\vskip 0.2in
\includegraphics[width=0.48\textwidth]{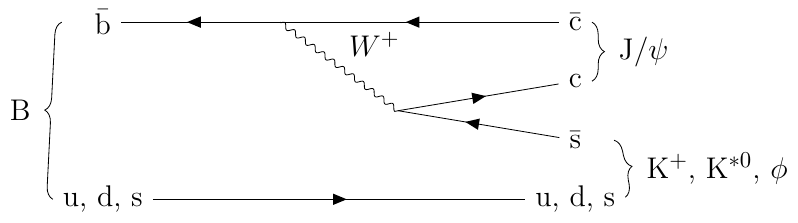}
\caption{Representative Feynman diagrams for the open-charm (upper) and charmonium (lower) decays of the \PBp, \PBz, and \PBzs mesons.}
\label{fig:feynman}
\end{figure}

In this paper, we extend the earlier CMS analysis~\cite{CMS:2022wkk}, which used 2018 data collected in $\Pp\Pp$ collisions at $\sqrt{s} = 13\TeV$ with dimuon triggers, by measuring $\fsfd$, $\fsfu$, and $\fdfu$ in a dedicated data set from the same year and collision energy, obtained with high-rate single-muon triggers designed to collect an unbiased sample of $10^{10}$ \PQb hadrons~\cite{CMS:2024zhe}. This data set, corresponding to an integrated luminosity of $41.6 \pm 1.0$\fbinv~\cite{LUM-18-001}, allows for a direct measurement of PFRs using the open-charm \PB meson decays \Bph, \Bzh, and \Bsh for the first time in CMS\@. Additionally, the relative PFR variables, \Rs and \Rsd, are measured simultaneously with charmonium decays (\Bpc, \Bzc, and \Bsc), where \Rs and \Rsd are defined as the ratio of efficiency-corrected yields of \PBzs to \PBp and \PBz mesons, respectively. We also derive the absolute normalization for \Rs and \Rsd by combining the results from the open-charm and charmonium decays, which can be used for extraction of the PFRs using charmonium decays. Another product of this analysis is the measurement of three \PB branching fraction ratios. Representative Feynman diagrams for the open-charm and charmonium decays are shown in Fig.~\ref{fig:feynman}. Tabulated data for this analysis can be found in the HEPData database~\cite{HEPData}.

The paper is organized as follows. After a brief description of the CMS detector in Section 2, we discuss the data and simulated event samples in Section 3, and outline the analysis strategy in Section 4. The event reconstruction and \PB meson candidate selection are described in Section 5, followed by the signal extraction in Section 6 and the efficiency determination in Section 7. Systematic uncertainties are discussed in Section 8, followed by the presentation of the results in Section 9. The paper is summarized in Section 10.

\section{The CMS detector}
\label{sec:CMS}
The central feature of the CMS apparatus is a superconducting solenoid of 6\unit{m} internal diameter, providing a magnetic field of 3.8\unit{T}. Within the solenoid volume are a silicon pixel and strip tracker, a lead tungstate crystal electromagnetic calorimeter, and a brass and scintillator hadron calorimeter, each composed of a barrel and two endcap sections. Forward calorimeters extend the pseudorapidity ($\eta$) coverage provided by the barrel and endcap detectors. Muons are measured in gas-ionization detectors embedded in the steel flux-return yoke outside the solenoid. 

During the 2018 LHC running period when the data used in this paper were recorded, the silicon tracker consisted of 1856 silicon pixels and 15\,148 silicon strip detector modules. Details on the pixel detector can be found in Ref.~\cite{Phase1Pixel}.
The silicon tracker measures charged particles within the pseudorapidity range $\abs{\eta} < 3$. For nonisolated particles of $1 < \pt < 10\GeV$, the track resolutions are typically 1.5\% in \pt and 20--75\mum in the transverse impact parameter~\cite{DP-2020-049}.

Muons are measured in the pseudorapidity range $\abs{\eta} < 2.4$, with detection planes made using three technologies: drift tubes, cathode strip chambers, and resistive-plate chambers. The single-muon trigger efficiency exceeds 90\% over the full pseudorapidity range, and the efficiency to reconstruct and identify muons is greater than 96\%. Matching muons to tracks measured in the silicon tracker results in a relative transverse momentum resolution, for muons with \pt up to 100\GeV, of 1\% in the barrel and 3\% in the endcaps~\cite{CMS:2018rym}.

Events of interest are selected using a two-tiered trigger system. The first level (L1), composed of custom hardware processors, uses information from the calorimeters and muon detectors to select events at a rate of around 100\unit{kHz} within a fixed latency of about 4\mus~\cite{CMS:2020cmk}. The second level, known as the high-level trigger (HLT), consists of a farm of processors running a version of the full event reconstruction software optimized for fast processing and reduces the event rate to a few\unit{kHz} before data storage~\cite{CMS:2016ngn}.

A more detailed description of the CMS detector, together with a definition of the coordinate system used and the relevant kinematic variables, can be found in Refs.~\cite{CMS:2008xjf,CMS:2023gfb}.

\section{Data and simulated samples}
\label{sec:data}
\subsection{The CMS \texorpdfstring{\PB}{B} parking campaign}
\label{sec:bparking}
During 2018 data taking period, CMS deployed a novel trigger and data processing strategy, referred to as ``\PB parking'', 
which enabled the collection and reconstruction of order $10^{10}$ unbiased \PQb hadron decays~\cite{CMS:2024zhe}. The trigger strategy used to collect these data exploited the fact that \PQb hadrons are predominantly produced in pairs at the LHC, and therefore one could use a semileptonic decay of one (``tag-side'') \PQb hadron to trigger on the event, while the other (``probe-side") \PQb hadron can decay arbitrarily, \ie, its decay is {\it unbiased\/} by the trigger.

The \PB parking trigger requires a muon with a relatively high \pt and a significant displacement from the $\Pp\Pp$ collision point. This triggering strategy is made possible by a relatively low rate of L1 single-muon triggers, combined with a gradual decrease of the L1 bandwidth and online computing resource usage as the instantaneous luminosity \lumi decreases over the LHC fill. Specifically, as \lumi decreases, the L1 and HLT trigger rates of the nominal CMS physics program also decrease. Therefore, the \PB parking trigger thresholds were tuned to use most of the available bandwidth and resources by switching on triggers with looser requirements as \lumi goes down.

Table~\ref{tab:trigger} summarizes the muon trigger requirements imposed at the L1 and HLT\@. The L1 trigger logic requires the presence of a muon within $\abs{\eta} < 1.5$ and imposes a variable minimum \pt requirement. Additionally, for the full pseudorapidity range of $\abs{\eta} < 2.4$, a single-muon L1 trigger with a fixed \pt threshold of 22\GeV is used. Since the muon displacement information is not available at L1, only \pt and \abs{\eta} thresholds are used to control the rate. At the HLT, the \pt threshold is refined, based on more precise momentum reconstruction compared to that at L1. Additionally, a requirement is imposed on the minimum two-dimensional muon track impact parameter significance \ipsigtrg, defined as the distance of closest approach of the track to the beamline in the plane transverse to the beams, divided by its uncertainty. Overall, this strategy leads to an L1 rate of up to about 30\unit{kHz} and an HLT output rate of up to 5.4\unit{kHz}.

The \PB parking data set corresponds to an integrated luminosity of $41.6 \pm 1.0$\fbinv~\cite{LUM-18-001}, about 20\% less than the integrated luminosity of 59.8\fbinv collected with the main physics triggers operating during the entire 2018 data taking.

\begin{table}[!htb]
    \centering
    \topcaption{Summary of the single-muon \PB parking triggers used in this analysis.}
    \begin{scotch}{lcccccc}
    \mystrut Peak \lumi	& L1 \ptm 		& HLT \ptm 	& \multirow{2}{*}{HLT \PGm \ipsigtrg}	& Peak HLT \\
                \lumunit       	& [\GeVns{}] 	& [\GeVns{}] 	&	& rate [kHz] \\
    \hline
    1.7 & ${>}12$ & ${>}12$ & ${>}6$ & 1.5 \\
    1.5 & ${>}10$ & ${>}9$  & ${>}6$ & 2.8 \\
    1.3 & ${>}9$  & ${>}9$  & ${>}5$ & 3.0 \\
    0.9 & ${>}7$  & ${>}7$  & ${>}4$ & 5.4 \\
    \end{scotch}
    \label{tab:trigger}
\end{table}

\subsection{Event simulation} 
\label{sec:sim}
Monte Carlo (MC) simulated samples are used to evaluate signal efficiencies, optimize the analysis, and model various background sources. Signal and background processes are generated using the \PYTHIA 8.230~\cite{Sjostrand:2014zea} package, including parton showering, fragmentation, and hadronization. The underlying event is also modeled with \PYTHIA using the CP5~\cite{CMS:2019csb} tune. The parton distribution functions are taken from the NNPDF3.1 set~\cite{NNPDF:2017mvq}. To account for the effects of additional interactions in the same or adjacent bunch crossings (pileup), simulated minimum bias events are added to the hard scattering, with the pileup vertex multiplicity matching that observed in the data. Samples are interfaced with \EVTGEN 1.3.0~\cite{Lange:2001uf}, which is used to simulate the decays of heavy hadrons. Final-state photon radiation is modeled with \PHOTOS 3.61~\cite{Barberio:1993qi}.

The following signal samples were generated with and without emulation of trigger conditions: 
\begin{itemize}
\itemsep 0pt
\item \Bph, 
\item \Bzh, 
\item \Bsh, 
\item \Bpc, 
\item \Bzc, and 
\item \Bsc. 
\end{itemize}
In addition, we simulated several background processes. In the open-charm channels, they are the Cabibbo-suppressed decays
\begin{itemize}
\itemsep 0pt
\item \BphK, 
\item \BzhK, and 
\item \BshK. 
\end{itemize}
In the charmonium channels, they are 
\begin{itemize}
\itemsep 0pt
\item $\PBp \!\to\! \JPsi(\MM)\PGpp$ and 
\item $\PGLzb \!\to\! \JPsi(\MM) \Pp\PKm$.
\end{itemize}

The CMS detector response is simulated using the \GEANTfour package~\cite{Agostinelli:2002hh}. Simulated samples are reconstructed with the same software packages as used for collision data.

\section{Analysis strategy}
\label{sec:strategy}
Measurements of the PFRs of \PB mesons are performed in the open-charm and charmonium decay modes. As independent branching fraction measurements of \PBzs meson decay modes with a precision better than about 20\% are not available, the only way to extract the PFR for the \PBzs meson is by using theoretical predictions. Precise theoretical calculations are currently available for the open-charm decays using the quantum chromodynamics factorization approach~\cite{Beneke:2000ry,Fleischer:2010ca,Huber:2016xod}. Since no reliable calculations exist for the charmonium modes, independent absolute measurements of \fsfd and \fsfu are impossible in this channel. Therefore, we utilize the measured PFRs in the open-charm channels and normalize the relative measurement of PFRs in the charmonium channels to those in order to extract the PFRs in the charmonium channels.  Measurements of \fdfu can be directly performed in both channels as they do not involve \PBzs decays.  To ensure that our measurement of \fdfu is a test of isospin invariance, it is important to avoid assuming isospin invariance.  The values of most $\PBz$ and $\PBp$ branching fractions are dominated by results from the \PQB factories, and often assume equal production of $\PBp\PBm$ and $\PBz\PABz$ pairs at the $\PGUP{4S}$ from isospin invariance.   There are indications that this assumption is not valid, and a recent evaluation has found $r^{\pm,0} \equiv \BR(\PGUP{4S}\!\to\! \PBp\PBm)/\BR(\PGUP{4S}\!\to\! \PBz\PABz) = 1.057 \pm 0.023$~\cite{Bernlochner:2023bad}.

\subsection{Open-charm measurement}
The most recent and precise theoretical calculations of the branching fractions of the \BzhKs and \Bshs decays~\cite{Bordone:2020gao} reveal a puzzling substantial discrepancy between the theory predictions and the experimental measurements in the individual decay channels. Nevertheless, the prediction for the ratio of these decays appears to be in good agreement with the experimental measurements~\cite{Bordone:2020gao}, so we will rely on this theoretical framework to extract the \fsfd ratio.

At the same time, the prediction for the branching fraction of the Cabibbo-favored \Bzhs decay is prone to a sizable additional uncertainty, as there is a nonfactorizable exchange diagram that contributes to this decay channel, which is absent in the case of the Cabibbo-suppressed \BzhKs decay. The current theoretical estimate of the impact of this diagram on the corresponding branching fraction is $0.966 \pm 0.056$~\cite{Fleischer:2010ca}, which contributes an additional ${\approx}6\%$ theoretical uncertainty. This presents an experimental challenge, given the lack of particle identification in CMS, as far as pion/kaon separation is concerned. Because of that, there is no difference in the experimental topology for the Cabibbo-suppressed and Cabibbo-favored decays, so the latter becomes a major background to the former, making it impractical to extract the \BzhKs signal in bins of \pt or rapidity of the \PBz meson. We overcome this difficulty by measuring the \Bzhs signal and then relating it to the theoretically clean \BzhKs signal by using the world-average of the ratio of these two branching fractions: $\BR(\BzhKs)/\BR(\Bzhs) = 0.0819 \pm 0.0020$~\cite{ParticleDataGroup:2024cfk}, which is known to a 2.4\% precision and is independent of the \fsfd value. That allows for the extraction of the \fsfd value in a theoretically clean fashion by using the much more copious Cabibbo-favored decay channel without introducing a large theoretical uncertainty due to the nonfactorizable exchange diagram contribution.

The formula used for the \fsfd measurement is as follows (using notations of Ref.~\cite{LHCb:2021qbv}):
\ifthenelse{\boolean{cms@external}}
{
\begin{widetext}
\begin{equation}
\begin{split}
\fsfd &= \frac{\BR(\Bzhs)}{\BR(\Bshs)} \frac{\BR(\Dmh)}{\BR(\Dsh)\BR(\phiKK)} \frac{\Ncorr(\Bsh)}{\Ncorr(\Bzh)} \\
&= \frac{\BR(\BzhKs)}{\BR(\Bshs)} \frac{\BR(\Bzhs)}{\BR(\BzhKs)} \frac{\BR(\Dmh)}{\BR(\Dsh)\BR(\phiKK)} \frac{\Ncorr(\Bsh)}{\Ncorr(\Bzh)} \\
&= \Phi_{\text{PS}} \frac{\abs{\Vus}^2}{\abs{\Vud}^2} \frac{f^2_\PK}{f^2_\PGp} \frac{\tBz}{\tBzs}\frac{1}{\Na\NF} \frac{\BR(\Bzhs)}{\BR(\BzhKs)} \frac{\BR(\Dmh)}{\BR(\Dsh)\BR(\phiKK)} \frac{\Ncorr(\Bsh)}{\Ncorr(\Bzh)},
\label{eq:fsfd}
\end{split}
\end{equation}
\end{widetext}
}
{
\begin{equation}
\begin{split}
  \fsfd &= \frac{\BR(\Bzhs)}{\BR(\Bshs)} \frac{\BR(\Dmh)}{\BR(\Dsh)\BR(\phiKK)} \\
& \qquad \frac{\Ncorr(\Bsh)}{\Ncorr(\Bzh)} \\
&= \frac{\BR(\BzhKs)}{\BR(\Bshs)} \frac{\BR(\Bzhs)}{\BR(\BzhKs)} \frac{\BR(\Dmh)}{\BR(\Dsh)\BR(\phiKK)} \\
&\qquad \times \frac{\Ncorr(\Bsh)}{\Ncorr(\Bzh)} \\
&= \Phi_{\text{PS}} \frac{\abs{\Vus}^2}{\abs{\Vud}^2} \frac{f^2_\PK}{f^2_\PGp} \frac{\tBz}{\tBzs}\frac{1}{\Na\NF} \frac{\BR(\Bzhs)}{\BR(\BzhKs)} \frac{\BR(\Dmh)}{\BR(\Dsh)\BR(\phiKK)} \\
&\qquad \times \frac{\Ncorr(\Bsh)}{\Ncorr(\Bzh)},
\label{eq:fsfd}
\end{split}
\end{equation}
}
where $\Phi_{\text{PS}} = 0.97074 \pm 0.00010$ is the ratio of the phase space factors for the two decays (given by inverting the expression on the last line of Eq.~(7) in Ref.~\cite{Bordone:2020gao}); \Vud and \Vus are the corresponding CKM matrix elements; \tBz and \tBzs are lifetimes of the \PBz and \PBzs mesons, respectively; \fpi and \fK are decay constants for the \PGp and \PK mesons; \Na and \NF are, respectively, the ratio of Wilson coefficients squared of the \PDp\PKm and \PDps\PGpm decays and the ratio of form-factors squared for the \BzhKs and \Bshs decays; and \Ncorr(\PX) is the efficiency- and acceptance-corrected event yield for process \PX. The last equality in Eq.~(\ref{eq:fsfd}) uses a theoretical calculation for the ratio of branching fractions of the \BzhKs to \Bshs decays from Ref.~\cite{Bordone:2020gao}.

For the \fdfu extraction, no theoretical calculations are needed and the following formula is used:
\ifthenelse{\boolean{cms@external}}
{
\begin{multline}
  \frac{\fd}{\fu} = \frac{\BR(\Bphs)}{\BR(\Bzhs)} \frac{\BR(\DzKpi)}{\BR(\DmKpipi)} \\
  \times \frac{\Ncorr(\Bzh)}{\Ncorr(\Bph)}.
\label{eq:fdfu}
\end{multline}
}
{
\begin{equation}
\frac{\fd}{\fu} = \frac{\BR(\Bphs)}{\BR(\Bzhs)} \frac{\BR(\DzKpi)}{\BR(\DmKpipi)} \frac{\Ncorr(\Bzh)}{\Ncorr(\Bph)}.
\label{eq:fdfu}
\end{equation}
}

Finally, \fsfu can be obtained from the product of Eqs.~(\ref{eq:fsfd}) and (\ref{eq:fdfu}) as:
\ifthenelse{\boolean{cms@external}}
{
\begin{widetext}
\begin{equation}
\fsfu = \Phi_{\text{PS}} \frac{\abs{\Vus}^2}{\abs{\Vud}^2} \frac{f^2_\PK}{f^2_\PGp} \frac{\tBz}{\tBzs}\frac{1}{\Na\NF} \frac{\BR(\Bphs)}{\BR(\BzhKs)} \frac{\BR(\Dzh)}{\BR(\Dsh)\BR(\phiKK)} \frac{\Ncorr(\Bsh)}{\Ncorr(\Bph)}.
\label{eq:fsfu}
\end{equation}
\end{widetext}
}
{
\begin{equation}
\begin{split}
\fsfu &= \Phi_{\text{PS}} \frac{\abs{\Vus}^2}{\abs{\Vud}^2} \frac{f^2_\PK}{f^2_\PGp} \frac{\tBz}{\tBzs}\frac{1}{\Na\NF} \frac{\BR(\Bphs)}{\BR(\BzhKs)} \frac{\BR(\Dzh)}{\BR(\Dsh)\BR(\phiKK)} \\
&\qquad \times \frac{\Ncorr(\Bsh)}{\Ncorr(\Bph)}.
\label{eq:fsfu}
\end{split}
\end{equation}
}

Utilizing the same measurements as in the average values of $\BR(\Bphs)$ and $\BR(\Bzhs)$ in Ref.~\cite{ParticleDataGroup:2024cfk}, but updating the charm branching fractions and converting to $r^{\pm,0} = 1.057 \pm 0.023$, we obtain a value of $\BR(\Bphs)/\BR(\Bzhs) = 1.763 \pm 0.066$ (not including the $r^{\pm,0}$ uncertainty).
The other values used in Eqs.~(\ref{eq:fsfd}--\ref{eq:fsfu}) are collected in Table~\ref{tab:inputsh}.

\begin{table*}[!htb]
    \centering
    \topcaption{Inputs for the open-charm PFR measurements.}
    \begin{scotch}{lcl}
    Input variable & Value	& Reference \\
    \hline
    $\Phi_{\text{PS}}$ & $0.97074 \pm 0.00010$ & PDG~\cite{ParticleDataGroup:2024cfk}; Ref.~\cite{Bordone:2020gao}, Eq.~(7) \\
    $\abs{V_{\PQu\PQs}}f_{\PK}/(\abs{V_{\PQu\PQd}}f_{\PGp})$ & $0.27683 \pm 0.00035$ & Ref.~\cite{DiCarlo:2019thl}, Eq.~(107) \\
    $\tau_{\PBz}/\tau_{\PBzs}$ & $1.0021 \pm 0.0034$ & HFLAV~\cite{HFLAV:2024,HFLAV:PDG2025} \\
    $1/\Na$ & $1.0048^{+0.0046}_{-0.0022}$ & Ref.~\cite{Bordone:2020gao}, Eq.~(20) \\
    $1/\NF$ & $1.002 \pm 0.042$ & Ref.~\cite{Bordone:2020gao}, Eq.~(16) \\
    $\BR(\BzhKs)/\BR(\Bzhs)$ & $0.0819 \pm 0.0020$ & PDG~\cite{ParticleDataGroup:2024cfk}\\
    $\BR(\Dmh)$ & $0.0938 \pm 0.0016$ & PDG~\cite{ParticleDataGroup:2024cfk}\\
    $\BR(\Dsh)\BR(\phiKK)$ & $0.0225 \pm 0.0005$ & PDG~\cite{ParticleDataGroup:2024cfk}\\
    $\BR(\Bphs)$ & $0.00461 \pm 0.00010$ & PDG~\cite{ParticleDataGroup:2024cfk}\\
    $\BR(\DzKpi)$ & $0.03945 \pm 0.00030$ & PDG~\cite{ParticleDataGroup:2024cfk}\\
     \end{scotch}
    \label{tab:inputsh}
\end{table*}

\subsection{Charmonium measurement}
In the charmonium channel, we use \Bpc, \Bzc, and \Bsc decays to measure the \Rs and \Rsd parameters, which are determined as follows:
\begin{equation}
\begin{split}
 \Rs & = \frac{\Ncorr(\Bsc)}{\Ncorr(\Bpc)} \\
       & =  \frac{\fs}{\fu} \frac{\BR(\Bscs)\BR(\phiKK)}{\BR(\Bpcs)}
\end{split}
\label{eq:Rs}
\end{equation}
and
\begin{equation}
\begin{split}
\Rsd & = \frac{\Ncorr(\Bsc)}{\Ncorr(\Bzc)} \\
 & = \frac{\fs}{\fd} \frac{\BR(\Bscs)\BR(\phiKK)}{\BR(\Bzcs)\BR(\KsKpi)}.
\end{split}
\label{eq:Rsd}
\end{equation}
The ratio \fdfu can be determined directly as:
\ifthenelse{\boolean{cms@external}}
{
\begin{multline}
\frac{\fd}{\fu} = \frac{\Ncorr(\Bzc)}{\Ncorr(\Bpc)} \\ \times \frac{\BR(\Bpcs)}{\BR(\Bzcs)\BR(\KsKpi)}.
\label{eq:fdfuc}
\end{multline}
}
{
\begin{equation}
\frac{\fd}{\fu} = \frac{\Ncorr(\Bzc)}{\Ncorr(\Bpc)} \frac{\BR(\Bpcs)}{\BR(\Bzcs)\BR(\KsKpi)}.
\label{eq:fdfuc}
\end{equation}
}
As for the open-charm states, the branching fractions used to obtain \Rs and \Rsd are from Ref.~\cite{ParticleDataGroup:2024cfk}.
For the determination of \fdfu, we use the same measurements as in the average values of $\BR(\Bpcs)$ and $\BR(\Bzcs)$ from Ref.~\cite{ParticleDataGroup:2024cfk}, but scaling the results to $r^{\pm,0} = 1.057 \pm 0.023$~\cite{Bernlochner:2023bad} to obtain $\BR(\Bpcs)/\BR(\Bzcs) = 0.769 \pm 0.033$ (not including the $r^{\pm,0}$ uncertainty).  The existing measurements of $\BR(\Bzcs)$ include one angular analysis, which results in a pure branching fraction measurement~\cite{Belle:2014nuw}, and the remainder, which select events in a mass region around the $\PKstn$ mass.  Using the model from Ref.~\cite{Belle:2014nuw} to estimate the $S$-wave contribution in the other measurements and scaling all of the measured branching fractions to correspond to the mass region used in our analysis results in a negligible change.

\section{Event reconstruction and \texorpdfstring{\PB}{B} candidate selection\label{sec:selection}}
Since the events used in this analysis are collected with the \PB parking triggers, we require that the muon which triggered the event readout (\mutrig) is reconstructed offline. The $z$ position of the point of closest approach of the triggering muon to the beamline is used to select the primary vertex (PV) of the hard interaction. Tracks are required to pass the ``high-purity" quality criteria~\cite{CMS:2010vmp,CMS:2017yfk}.

\subsection{Open-charm analysis}
The \PB candidate selection relies on the probe side of the event. We require the tracks used to build the \PB candidate to be separated from the triggering muon by $\Delta R(\text{track},\mutrig) > 0.4$, where $\Delta R = \sqrt{\smash[b]{(\Delta\eta)^2 + (\Delta\phi)^2}}$, and $\phi$ is the azimuthal angle in radians. To reduce pileup, the tracks are also required to originate from the vicinity of the PV, with $\Delta z(\text{track},\text{PV}) < 0.5\unit{cm}$. Tracks are further selected to have $\abs{\eta} < 2.4$ and $\pt > 1\GeV$. Since the \PB mesons result in displaced secondary vertices (SVs) due to their sizable decay distance, we only consider tracks with $\ipsig > 1$, where $\ipsig$ is defined as the transverse distance of closest approach of a track to the PV divided by its uncertainty.

Depending on the final state, \PB candidates have either three (\PBp) or four (\PBz and \PBzs) tracks originating from the final-state particles. We create B candidates out of all combinations of pion and kaon mass assignments to these tracks. 
Since the \PD meson from the \PB candidate decay will be displaced from the \PB meson decay SV, we start the \PB candidate reconstruction by finding a tertiary vertex (TV) of a decay of the corresponding \PD meson. The \Dzh, \DmKpipi, and  $\PDms \!\to\! \PGpm\PKp\PKm$ candidates are reconstructed by selecting the correct number of tracks with the correct charge and an invariant mass within 20, 20, and 15\MeV of the world-average \PD meson masses $m(\PDz)$, $m(\PDm)$, and $m(\PDms)$~\cite{ParticleDataGroup:2024cfk}, respectively. In all three decay channels, the highest \pt (leading) track of the pair or triplet is required to have $\pt > 1.5\GeV$. The tracks of the \PD candidate are then fitted to a TV, and the \PD candidate is accepted if the fit probability exceeds 0.01 and the significance of the displacement of the TV from the PV in the plane transverse to the beams $S_{xy,\text{TV}}$, defined as the distance between the two vertices in the $x$-$y$ plane divided by its uncertainty, exceeds 5. The SV position is then obtained by fitting the remaining track with $\pt > 1.5\GeV$ and correct charge, and the \PD candidate momentum vector to a common vertex. The \PB candidate is then accepted if the significance of the transverse displacement of the SV from the PV, $S_{xy,\text{SV}}$, defined in a similar way as $S_{xy,\text{TV}}$, exceeds 7, and the cosine of the angle between the line connecting the PV and SV and the \PB candidate momentum vector, $\cos\alpha_\PB$, exceeds 0.999.

In the \PBzs channel, an additional requirement on the invariant mass of the $\PKp\PKm$ pair to be within 10\MeV of the world-average $\Pphi$ mass~\cite{ParticleDataGroup:2024cfk} is applied. This is crucial to remove the copious background from $\PBz \!\to\! \PGpp\PDm(\PKp\PGpm\PGpm)$ decays with one of the pions being misassigned a kaon mass, which results in the shift of the four-track invariant mass peak into the \PBzs region. The narrowness of the $\Pphi$ resonance ensures that this background is negligible (rejection power of over 200) after this extra selection, while retaining about 85\% of the signal. The contribution from the $\PKp\PKm$ continuum in the \PDms decay mode within the \Pphi meson mass window can be estimated from a Dalitz plot analysis of this decay mode by BaBar~\cite{BaBar:2010wqe}. The parameterization of the invariant mass distribution for the \Pphi and nonresonant components, as described in Ref.~\cite{BaBar:2010wqe}, is applied to fit the $\PKp\PKm$ invariant mass distribution in data. Based on the fit results, the nonresonant contribution in the \PBzs channel, which amounts to about 5\%, is subsequently subtracted using the $\mathit{_sPlot}$ technique~\cite{Pivk:2004ty}.

The summary of the preselection requirements to identify \PB candidates is given in Table~\ref{tab:selhad}. While multiple \PB candidates per event are allowed, $\approx$98\% of events have only a single candidate per event.

\begin{table*}[bht]
\renewcommand{\arraystretch}{1.2}
\topcaption{Summary of the preselection requirements for \PB candidates in the open-charm analysis. The $\NA$ entries indicate that the selection does not apply to this channel.}
\label{tab:selhad}
\centering
\begin{scotch}{lccc}
 Variable & \multicolumn{3}{c}{Selection} \\
  & \PBp & \PBz & \PBzs \\
\hline
\multicolumn{4}{c}{General track selection} \\
\hline
Track \pt 			& \multicolumn{3}{c}{${>}1\GeV$}\\
Track $\abs{\eta}$	& \multicolumn{3}{c}{${>}2.4$} \\
$\Delta z(\text{track},\mutrig)$ & \multicolumn{3}{c}{${<}0.5\unit{cm}$} \\
Track \ipsig & \multicolumn{3}{c}{$>1$} \\ [0.5ex]
\hline
\multicolumn{4}{c}{\PD candidate selection} \\
\hline
\PD candidate leading-track \pt & \multicolumn{3}{c}{${>}1.5\GeV$} \\
TV fit probability & \multicolumn{3}{c}{$>0.01$} \\
$S_{xy,\text{TV}}$ & \multicolumn{3}{c}{$>5$} \\
\PD candidate mass window & $\pm 20\MeV$ & $\pm 20\MeV$ & $\pm 15\MeV$ \\
$\abs{M(\PKp\PKm) - m(\Pphi)}$  &\NA & \NA & ${<}10\MeV$ \\
\hline
\multicolumn{4}{c}{\PB candidate selection} \\
\hline
\pt of the remaining \PB candidate track &  \multicolumn{3}{c}{${>}1.5\GeV$} \\
$S_{xy,\text{SV}}$ & \multicolumn{3}{c}{${>}7$} \\
$\cos\alpha_\PB$  & \multicolumn{3}{c}{${>}0.999$} \\
\end{scotch}
\end{table*}

To further improve signal purity, we perform a multivariate analysis (MVA) based on boosted decision trees (BDTs). One BDT is trained for each of the three channels. The preselection described above improves the BDT performance, as it reduces the bulk of backgrounds with easily distinguishable topology, thus allowing for the BDT to optimize the rejection against the remaining backgrounds, which are topologically much closer to the signal.

The BDTs are trained on simulated signal events with the \PB candidate mass in a signal region (SR), centered on the corresponding \PB meson mass, against the background, which is taken from the upper sideband (SB) of the data. This SB is dominated by the combinatorial background, mainly composed of combinations of a \PD meson and a random track, similar to the background in the SR. In contrast, the lower sideband is dominated by partially reconstructed \PB decays, which do not populate the SR and are therefore not used for the BDT training. Signal samples are weighted so that their normalization corresponds to the signal yield in data after the preselection. The SR is defined using the invariant mass of the \PB candidate as 5.20--5.35\GeV (5.30--5.45\GeV), while the SB is defined as 5.40--5.55\GeV (5.50--5.65\GeV) in the \Bphs and \Bzhs (\Bshs) channels. The following variables are used as inputs to the BDTs: \ipsig of all three or four tracks of the \PB candidate; invariant mass of the \PD candidate (and of the \Pphi candidate in the \Bshs channel) before the TV fit; fit probabilities for the SV and TV; $\cos\alpha_\PB$; and displacement significances in the $x$-$y$ plane of the SV with respect to the PV and of the TV with respect to the SV. The BDTs are trained using the \textsc{XGBoost}~\cite{Chen:2016btl} 0.72 package, with gradient boosting. Hyperparameters are optimized via a Bayesian optimization procedure using the \textsc{scikit-optimize}~\cite{Pedregosa:2011ork} 0.9.0 package. The cutoffs on the BDT discriminants were optimized using the $S/\sqrt{S+B}$ figure of merit (FOM), where $S$ and $B$ denote signal and background yields, respectively. The FOM after the BDT selection is increased by about a factor of 2 with respect to the preselection. For an optimized cutoff-based analysis using the same input variables and FOM as the BDT, the FOM is about 20\% lower than for the BDT selection. Consequently, the BDT analysis is chosen as the default approach for the signal extraction and the PFR measurement. It was checked that the output BDT variables in data obtained from background-subtracted \PB candidates after the BDT selection are well described by the signal simulation. The BDT was also checked for a potential mass bias (``sculpting"), and none was found.

\subsection{Charmonium analysis}
In the charmonium channel, given the higher purity of the signal, no significant improvement is expected from an MVA, so a cutoff-based analysis is used instead. The selection of the \PB candidates starts with a requirement of an opposite-sign (OS) muon pair with 
each muon passing the {\it soft\/} identification requirements~\cite{CMS:2018rym} and having a distance of closest approach to the PV in the $z$ direction of $\Delta z <1\unit{cm}$. Each muon is required to have $\pt > 3.5 \GeV$ in the barrel ($\abs{\eta} < 1.1$), $\pt > 1.5 \GeV$ in the endcap ($1.5 < \abs{\eta} < 2.4$), and $\pt > 3.5 + \frac{(1.5 - 3.5)}{(1.5 - 1.1)}(\abs{\eta} - 1.1) \GeV$ in the transition region. The variable muon \pt threshold reflects the muon detector acceptance in the barrel and endcap, which reduces the sensitivity to potential mismodeling at the edges of the detector acceptance. The muon pair is then fit to a common vertex, with its invariant mass required to be within $\pm 150 \MeV$ of the nominal $\JPsi$ meson mass~\cite{ParticleDataGroup:2024cfk}. For the \Bzcs (\Bscs) channel, we require two OS tracks with leading track $\pt > 0.85\GeV$, subleading track $\pt > 0.6\GeV$, $\abs{\eta} < 2.4$, and $\Delta z < 1\unit{cm}$ fitted to a common vertex, with the invariant mass consistent with the world-average \PKstn (\Pphi) mass~\cite{ParticleDataGroup:2024cfk} under the assignment of one kaon and one pion (two kaon) masses to the two tracks: 0.842--0.942\GeV (1.01--1.03\GeV). In the \PKstn case, both possible mass assignment permutations are tried, and in case both pass the invariant mass requirement, only the combination with the mass closest to the nominal \PKstn mass~\cite{ParticleDataGroup:2024cfk} is retained. For the \Bpcs channel, a single track with $\pt > 0.85\GeV$, $\abs{\eta} < 2.4$, and $\Delta z < 1\unit{cm}$ is required. The pair of muons and a track (\Bpcs channel) or a pair of tracks (\Bzcs and \Bscs channels) are fit to a common SV, with the \JPsi meson mass constraint on the pair of muons, and the SV $\chi^2$ fit probability is required to exceed 0.07. We further require $\cos\alpha_\PB > 0.997$, $S_{xy,\text{SV}} > 4$, and the \PB candidate mass within 4.5--6.0\GeV.

\begin{table*}[bht]
\renewcommand{\arraystretch}{1.2}
\topcaption{Summary of the selection requirements for \PB candidates in the charmonium analysis. The $\NA$ entries indicate that the selection does not apply to this channel.}
\label{tab:selcc}
\centering
\begin{scotch}{lccc}
 Variable & \multicolumn{3}{c}{Selection} \\
  & \PBp & \PBz & \PBzs \\
\hline
\multicolumn{4}{c}{General track selection} \\
\hline
Leading (subleading) track \pt 	& $>0.85\GeV$ & $>0.85$ (0.6)\GeV & $>0.85$ (0.6)\GeV \\
Track $\abs{\eta}$	& \multicolumn{3}{c}{${>}2.4$} \\
$\Delta z(\text{track},\text{PV})$ & \multicolumn{3}{c}{${<}1\unit{cm}$} \\
\hline
\multicolumn{4}{c}{Muon candidate selection} \\
\hline
Muon \pt & \multicolumn{3}{c}{$\pt>\left\{
\begin{array}{l}
3.5\GeV : \abs{\eta}<1.1 \\
3.5 + \frac{1.5-3.5}{1.5-1.1} (\abs{\eta}-1.1)\GeV : 1.1 < \abs{\eta} < 1.5 \\
1.5\GeV : \abs{\eta} > 1.5 \\
\end{array}\right.$} \\
Muon identification & \multicolumn{3}{c}{Soft~ID~\cite{CMS:2018rym}} \\
$\abs{M(\mu\mu) - m(\JPsi)}$ & \multicolumn{3}{c}{$<150\MeV$} \\
\hline
\multicolumn{4}{c}{\PB candidate selection} \\
\hline
SV fit probability & \multicolumn{3}{c}{$>0.07$} \\
$S_{xy,\text{SV}}$ & \multicolumn{3}{c}{$>4$} \\
$\cos\alpha_\PB$  & \multicolumn{3}{c}{$>0.997$} \\
\hline
\multicolumn{4}{c}{Intermediate resonance masses} \\
\hline
$\abs{M(\PKp\PKm) - m(\Pphi)}$ & \NA & ${>}10\MeV$ & ${<}10\MeV$ \\
$\abs{M(\PKp\PGpm) - m(\PKstn)}$ & \NA & ${<}50\MeV$, & ${>}50\MeV$, \\
 & & closest permutation & both permutations \\
\end{scotch}
\end{table*}

The summary of the selection requirements is presented in Table~\ref{tab:selcc}. These optimized selections are applied for all three charmonium channels, with the exception of the mass windows for the \Pphi selection and the \PKstn veto: they are inverted for the \Bzcs channel and are not used in the \Bpcs channel. The \Pphi (\PKstn) veto ensures that there is no overlap between the \Bzcs (\Bscs) candidates and the \Bscs (\Bzcs) candidates.

In the charmonium channel, we further split the analysis into the ``tag-side'' and ``probe-side'' categories. The tag-side category requires one of the muons from a \JPsi meson decay to trigger the event, while for the probe-side category, an additional muon that does not belong to the \JPsi candidate is required to trigger the event. If one event falls into both the tag-side and probe-side categories, it is classified as a probe side; thus, the two categories are mutually exclusive. Given that a charmonium analysis based on a significantly larger data set obtained with dimuon triggers has already been published by CMS~\cite{CMS:2022wkk} to test the kinematic dependence of PFRs, the goal of the present charmonium analysis is to examine the dependence of PFRs with minimal trigger bias, which is achievable by using the probe side. The translation factors between the \Rs (\Rsd) and \fsfu (\fsfd) variables are also measured on the probe side, as in the open-charm analysis, to minimize systematic uncertainties. The tag-side category is primarily used to verify the consistency of the present analysis with previous measurements.

\section{Signal extraction}
\label{sec:signal}
Since the main goal of the analysis is to measure \fsfd as a function of \pt and $\absy$ of the \PB candidates, we extract the signal by performing a binned maximum likelihood fit to the \PB candidate mass spectrum for each \pt or \absy bin with a sum of several functions describing signal and background components.

\subsection{Open-charm analysis}
For the open-charm analysis, there are seven $\pt(\PB)$ bins with the lower boundaries of 8, 10, 13, 18, 23, 28, and 33\GeV, with the last bin extending up to 60\GeV, and six bins in $\abs{y(\PB)}$ with the first five covering 0--1.25 in 0.25-wide bins and the last bin covering 1.25--2.25.

To improve the \PB candidate mass resolution, we use the \PD meson mass correction defined as follows for the \Bph, \Bzh, and \Bsh channels, respectively:
\begin{align*}
M(\PBp) & = M(\PKp\PGpm\PGpp) - M(\PKp\PGpm) + m(\PDz),\\
M(\PBz) & = M(\PKp\PGpm\PGpm\PGpp) - M(\PKp\PGpm\PGpm) + m(\PDm),\\
M(\PBzs) & = M(\PKp\PKm\PGpm\PGpp) - M(\PKp\PKm\PGpm) + m(\PDms).
\end{align*}

We use a sum of two Gaussian functions with a common mean (double-Gaussian function) to describe the signal template, with the mean, two widths, and the relative normalization parameter constrained via a Gaussian probability density function to the values and the uncertainties obtained from a fit to simulated distributions.

The Cabibbo-suppressed background where the pion is replaced with a kaon is described with the Johnson function~\cite{Johnson}, given by:
\begin{equation*}
\frac{\delta}{\lambda\sqrt{2\pi}} \frac{1}{\sqrt{1 + \left( \frac{x-\mu}{\lambda} \right)^2}} e^{-\frac{1}{2} \left(\gamma + \delta \sinh^{-1}\left(\frac{x-\mu}{\lambda}\right) \right)^2},
\end{equation*}
where $\mu$ and $\lambda$ are the mean and the width parameters, while $\gamma$ and $\delta$ are the parameters describing the tails. All of these parameters are also Gaussian-constrained to the values obtained from a fit to simulation. The normalization of this background is fixed relative to the Cabibbo-favored signal normalization. Specifically, we calculate the ratio of the signal efficiencies for the Cabibbo-suppressed and Cabibbo-favored channels in each bin and multiply them by the ratio of the corresponding branching fractions. The ratio of the efficiencies is found to be flat in both \pt and \absy, so we use a single number per channel: $\epsilon(\PK/\PGp) = 1.040^{+0.014}_{-0.021}$, $0.994^{+0.036}_{-0.040}$, and $1.007^{+0.079}_{-0.053}$ for the \Bphs, \Bzhs, and \Bshs channels, respectively. While these ratios are all consistent with unity, we nevertheless propagate them into the ratio of the corresponding branching fractions~\cite{ParticleDataGroup:2024cfk}, which yields the following normalization ratios between the expected yields in the Cabibbo-suppressed and Cabibbo-favored channels: $R_{\PK/\PGp} = 0.0806 \pm 0.0023$, $0.0817 \pm 0.0056$, and $0.0760 \pm 0.0064$, respectively. In the fits, both the relative normalization of the Cabibbo-suppressed background and the parameters of the Johnson function are constrained within their Gaussian uncertainties using the above $R_{\PK/\PGp}$ numbers and the shape parameters ($\mu$, $\lambda$, $\gamma$, $\delta$) determined from the fits to simulated samples.

All the open-charm channels have contributions from partially reconstructed backgrounds stemming from processes similar to the signal one, but with an additional pion that is not reconstructed, such as $\PB_{(\PQs)} \!\to\! \PGp\PD^{*}_{(\PQs)}$ and $\PB_{(\PQs)} \!\to\! \rho\PD_{(\PQs)}$. This background is small and also shifted to lower masses, and it is described via an error function with the shape parameters and normalization being free parameters of the fit.

\begin{figure}[bhtp]
\centering
\includegraphics[width=\figwid]{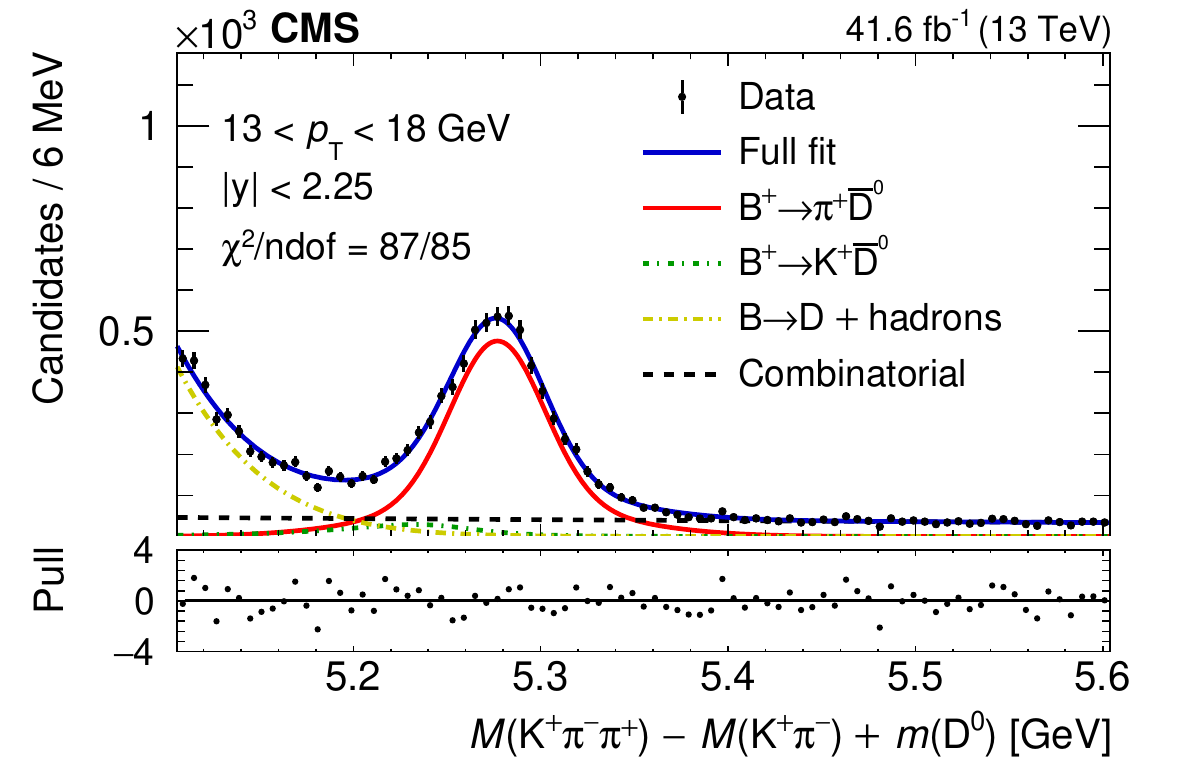}
\includegraphics[width=\figwid]{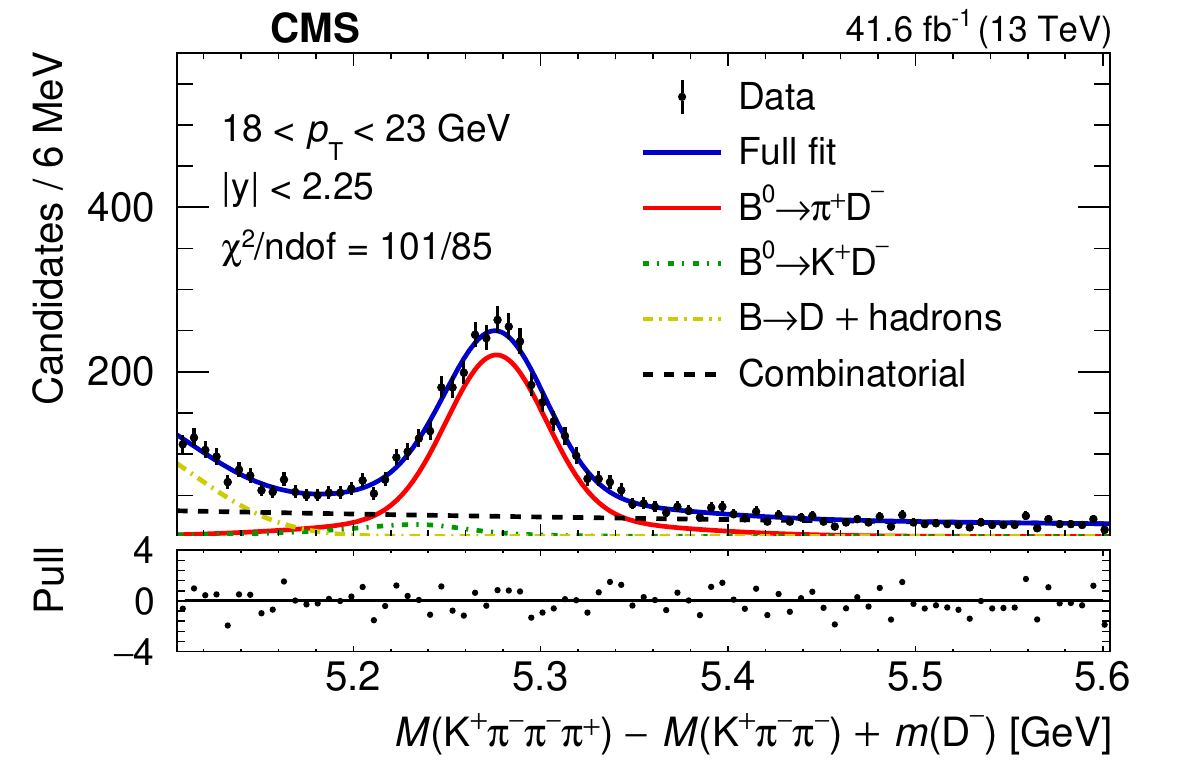}
\includegraphics[width=\figwid]{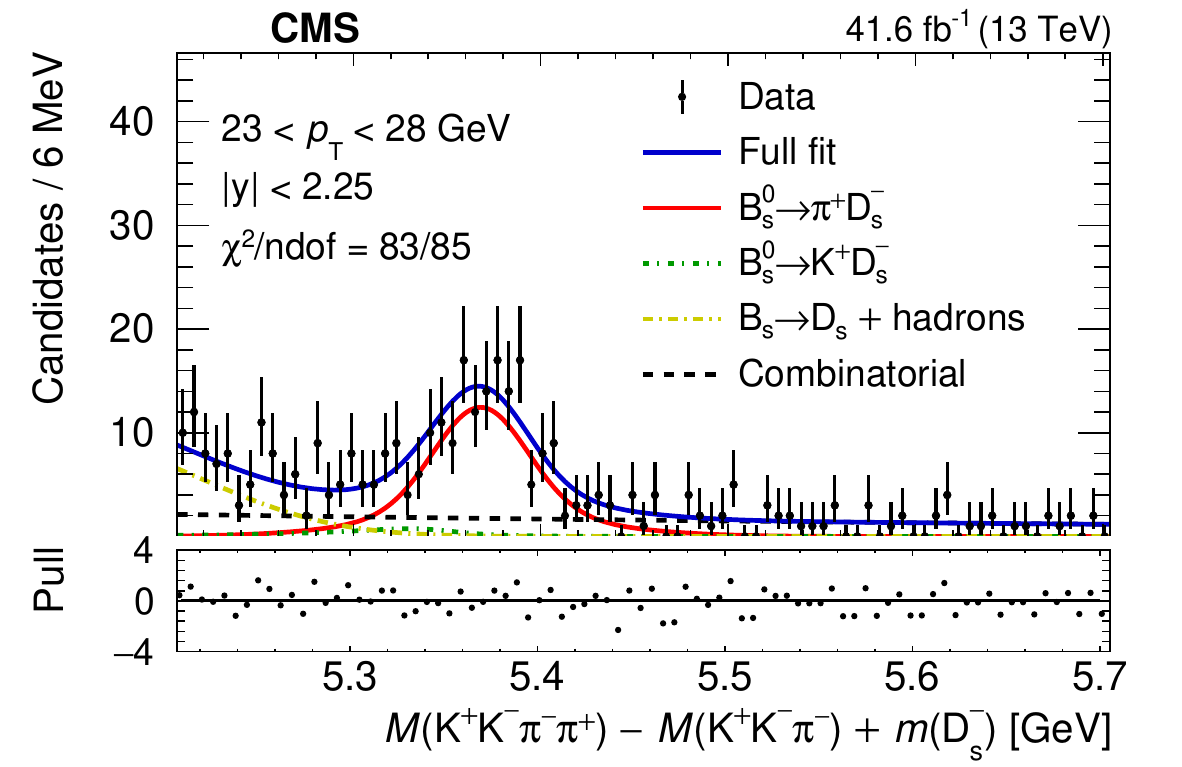}
\caption{Examples of fits to the \PB candidate mass distribution (corrected by the \PD meson mass) used to extract the signal in the  open-charm channels. The upper, middle, and lower plots correspond to the \PBp, \PBz, and \PBzs decays, respectively. The $\chi^2$ per degree of freedom ($\chi^2/\mathrm{ndof}$) of the fit is shown in each plot. The lower panels illustrate the pull, defined as the deviation of data from the fit function, normalized by the statistical uncertainty in each bin.}
\label{fig:fit_had}
\end{figure}

The combinatorial background is present in all channels and is modeled using an exponential function with both the normalization and slope as free parameters of the fit. Typical fits in the open-charm channels are presented in Fig.~\ref{fig:fit_had}.

\subsection{Charmonium analysis}
For the charmonium analysis, the binning and the \absy range are different for the tag- and probe-side categories. For the tag-side analysis, there are 13 $\pt(\PB)$ bins: five 1\GeV wide bins spanning the 11--16\GeV range, two 2\GeV wide bins spanning the 16--20\GeV range, three 3\GeV wide bins spanning the 20--29\GeV range, and three high-\pt bins: 29--34, 34--45, and 45--60 \GeV. For the probe-side analysis, we have fewer bins as there are fewer probe-side events. There are 6 $\pt(\PB)$ bins: five bins with a width of 5\GeV each, covering the 8--33\GeV range, and a sixth bin covering the 33--50\GeV range. The tag-side analysis has 6 \absy bins of 0.25 width each, spanning the 0--1.5 range, while the probe-side analysis has 9 bins of the same width, spanning the 0--2.25 range. The reason the tag-side analysis is more restricted in \absy is the requirement of \mutrig to have  $\abs{\eta}<1.5$ for the low-threshold \PB parking triggers. Finer binning in \abs{y} for the tag side is not considered, as the absence of \abs{y} dependence of PFRs has been well-verified in previous studies by both LHCb~\cite{LHCb:2021qbv} and CMS~\cite{CMS:2022wkk}.

Since the kinematic fit already constrains the \MM pair mass to the \JPsi meson mass, the invariant mass of the \PB candidate is directly used for signal extraction instead of correcting it with the \JPsi meson mass.

In the charmonium channel, signal shapes are modeled with Johnson functions, with the means and widths as free parameters of the fit, while the $\gamma$ and $\delta$ tail parameters are constrained by a fit to simulated signal events.

The Cabibbo-suppressed background from $\PBp \!\to\! \JPsi(\MM)\PGpp$ decays only enters in the \Bpcs channel. It is described by a sum of Crystal Ball (CB)~\cite{Oreglia} and Gaussian functions with a common mean. Given that the contribution from this background is small, we fix its relative normalization to $R_{\PK/\PGp} = 0.0385$~\cite{ParticleDataGroup:2024cfk}, and the tail parameters for the CB function, as well as the common mean, and the widths of both the CB and Gaussian functions, to the values obtained from a fit to a sample of simulated events.

The \Bpcs channel of the charmonium analysis has contributions from partially reconstructed backgrounds from the 
$\PB \!\to\! \JPsi\PKst$ process with one of the two tracks from the $\PKstn \!\to\! \PKp\PGpm$ decay, or a neutral meson in the
$\PKstp \!\to\! \PKp\PGpz$ or $\PKstp \!\to\! \PKz\PGpp$ decay lost. As in the open-charm channels, this background is small and shifted to lower masses, so it is similarly described by an error function with the shape parameters and normalization as free parameters of the fit.

The combinatorial background is modeled across all channels using an exponential function with both the normalization and slope as free parameters in the fit.

Finally, for the \Bzcs channel, the $\PGp/\PK$-swap component (\ie, the contribution to the signal shape when the swapped (incorrect) mass assignments of the $\PGp$ and $\PK$ tracks is used) is considered a signal component and described via a Johnson function with the shape parameters and the relative normalization to the signal (\ie, the nonswapped case) yield fixed from a fit to a sample of simulated signal events.

\begin{figure}[bhtp]
\centering
\includegraphics[width=\figwid]{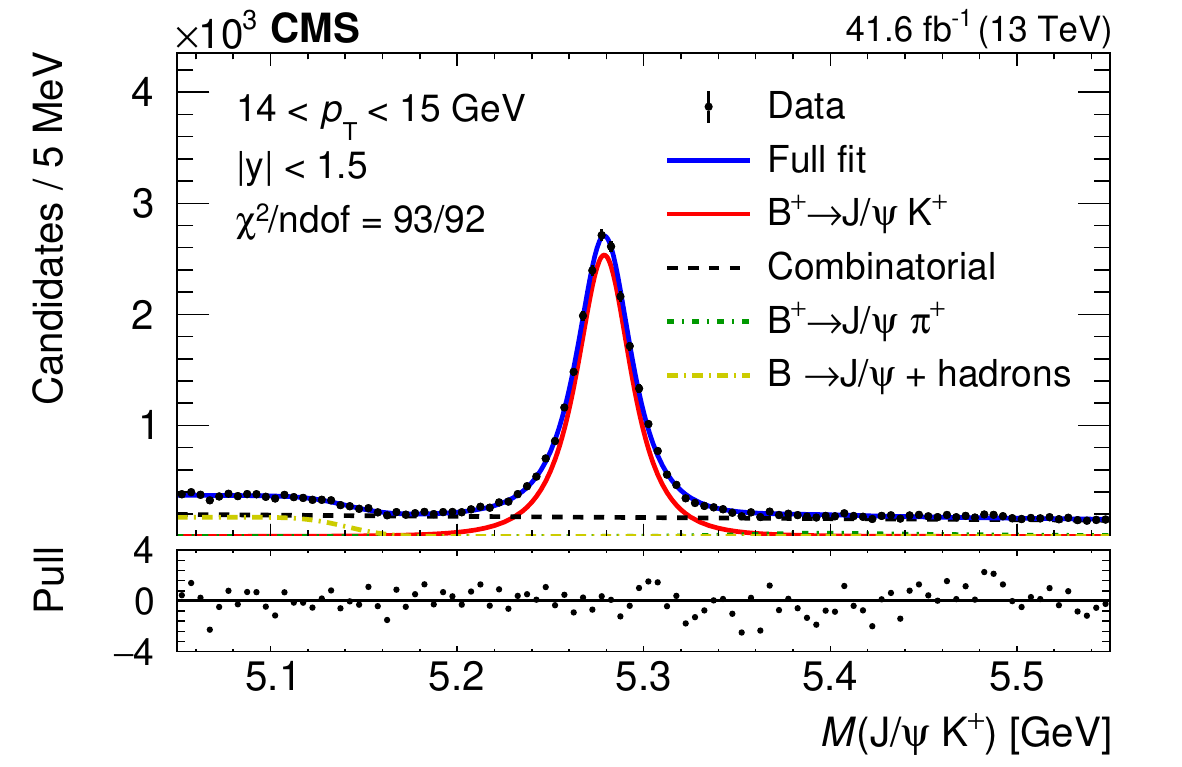}
\includegraphics[width=\figwid]{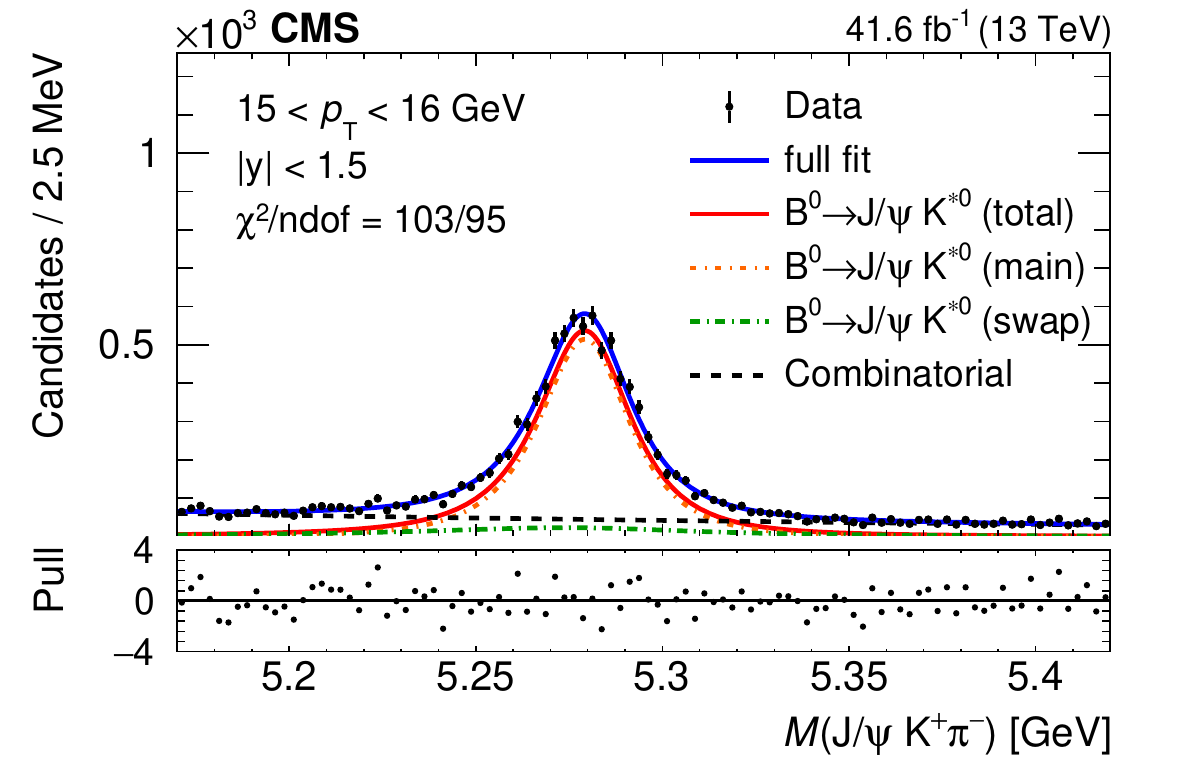}
\includegraphics[width=\figwid]{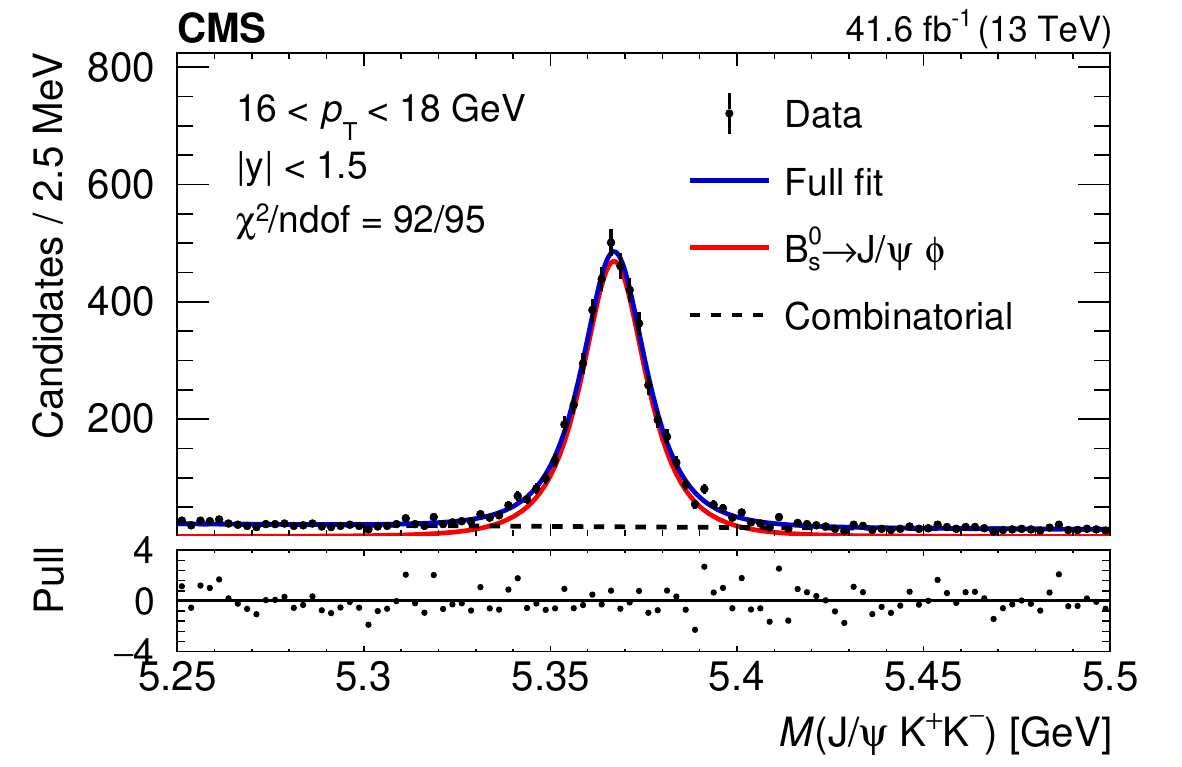}
\caption{Examples of fits to the \PB candidate mass distribution used to extract the signal in the charmonium channels of the tag-side analysis. The upper, middle, and lower plots correspond to the \PBp, \PBz, and \PBzs decays, respectively. The $\chi^2$ per degree of freedom ($\chi^2/\mathrm{ndof}$) of the fit is shown in each plot. The lower panels illustrate the pull, defined as the deviation of data from the fit function, normalized by the statistical uncertainty in each bin.}
\label{fig:fit_charm_tag}
\end{figure}

\begin{figure}[bhtp]
\centering
\includegraphics[width=\figwid]{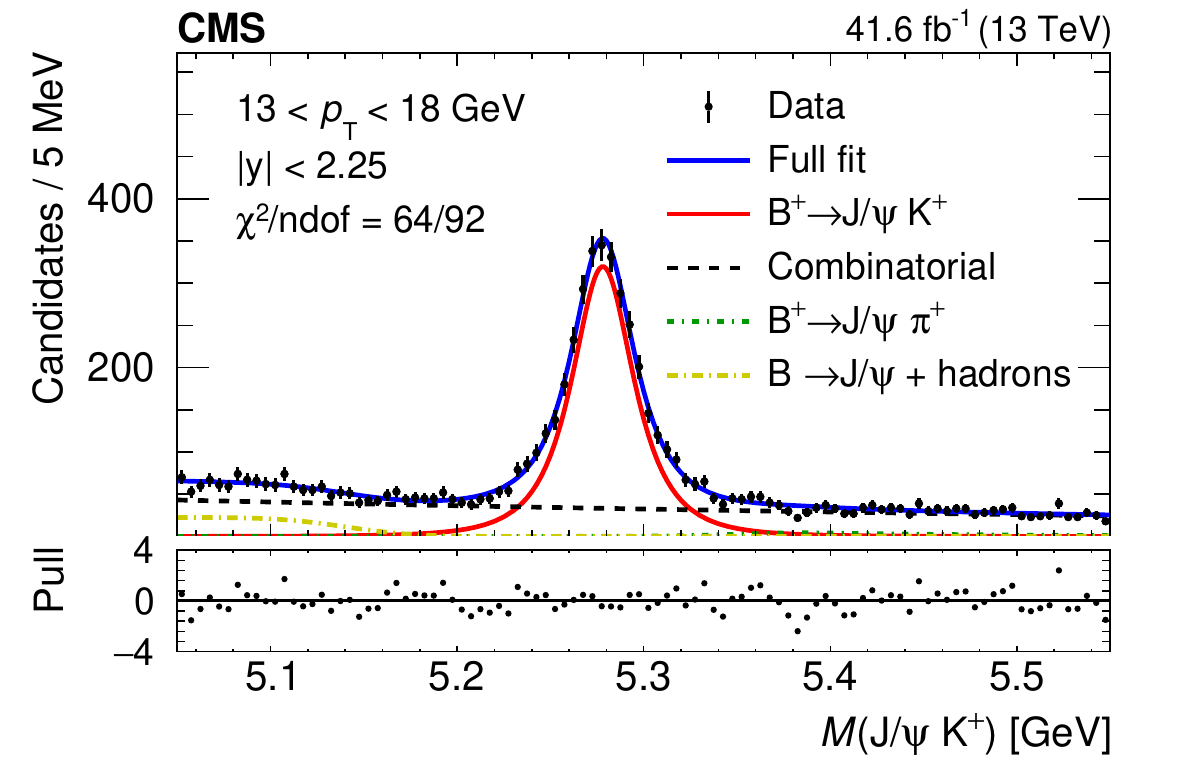}
\includegraphics[width=\figwid]{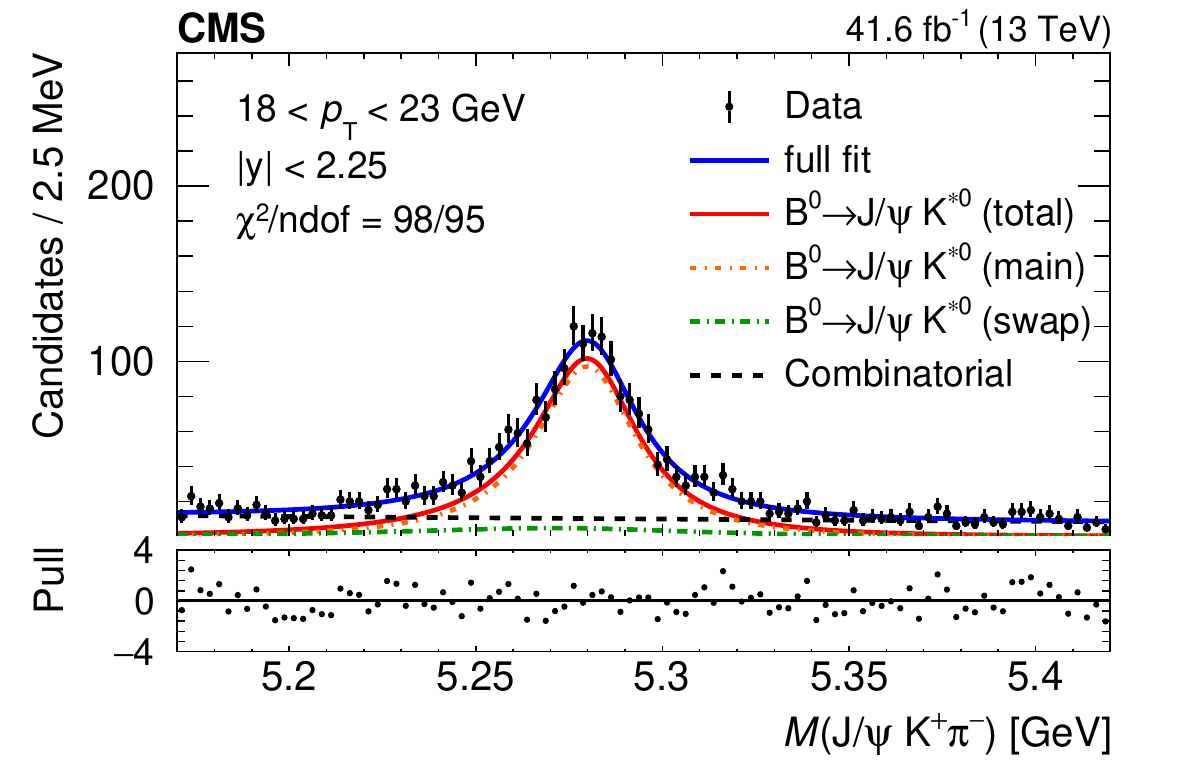}
\includegraphics[width=\figwid]{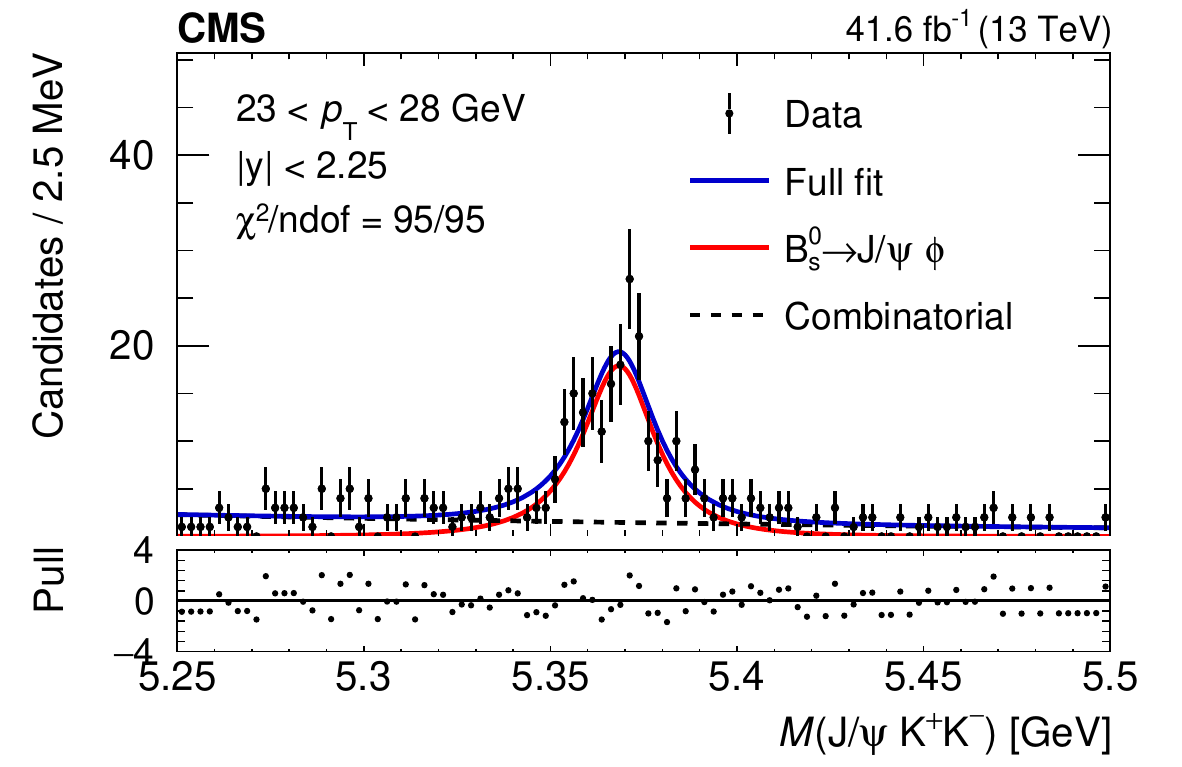}
\caption{Examples of fits to the \PB candidate mass distribution used to extract the signal in the charmonium channels of the probe-side analysis. The upper, middle, and lower plots correspond to the \PBp, \PBz, and \PBzs decays, respectively. The $\chi^2$ per degree of freedom ($\chi^2/\mathrm{ndof}$) of the fit is shown in each plot. The lower panels illustrate the pull, defined as the deviation of data from the fit function, normalized by the statistical uncertainty in each bin.}
\label{fig:fit_charm_probe}
\end{figure}

Figures~\ref{fig:fit_charm_tag} and~\ref{fig:fit_charm_probe} show typical fits in the charmonium channels for the tag-side and probe-side analyses, respectively. Table~\ref{tab:fit} summarizes the functions used to describe the signal and background components in each channel. 

\begin{table*}[bhtp]
\topcaption{Fit functions used in the open-charm and charmonium analyses. The $\NA$ entries indicate that the background does not apply to this channel.}
\label{tab:fit}
\centering
\begin{scotch}{lcccc}
 	     & \multicolumn{2}{c}{Open-charm analysis} &  \multicolumn{2}{c}{Charmonium analysis} \\
 Source & Function & Constraint & Function & Constraint \\
 \hline
 Signal & Double-Gaussian & Gaussian  &  Johnson & $\mu$, $\lambda$: free;  \\
	   & & & & $\delta$, $\gamma$: Gaussian \\
Cabibbo-supp. bkg. & Johnson & Gaussian & Crystal Ball & Fixed \\
	   & & & + Gaussian& \\
Partially rec. bkg. & Error function & Free & Error function & Free \\
Combinatorial & Exponential & Free & Exponential & Free \\
$\PGp/\PK$ swap signal & \multicolumn{2}{c}{\NA} & Johnson & Fixed \\
\end{scotch}
\end{table*}

\section{Efficiency determination}
\label{sec:efficiency}
As shown in Eqs.~(\ref{eq:fsfd}--\ref{eq:fdfuc}), the ratios of efficiency-corrected yields ($N_{\text{corr}}$) are used to obtain \Rs, \Rsd, and the PFRs. Therefore, only the corresponding efficiency ratios, $\epsilon_{\PBp}/\epsilon_{\PBzs}$, $\epsilon_{\PBz}/\epsilon_{\PBzs}$, and $\epsilon_{\PBz}/\epsilon_{\PBp}$, are required to correct the signal event yields in this analysis. These efficiency ratios are determined from samples of simulated events and reflect the trigger, reconstruction, and selection processes, as well as the detector acceptance.

While the efficiency ratios of the open-charm analysis and the probe-side charmonium analysis are minimally affected by the triggers, the tag-side charmonium analysis relies on a trigger muon. The difference in the trigger response between simulation and data on the tag side is corrected via muon trigger scale factors (SFs). The SFs for the \PB parking triggers used in this analysis are measured using the ``tag-and-probe'' method~\cite{CMS:2010svw} based on the $\JPsi\!\to\!\MM$ decays across different muon \pt and $\ipsig$ ranges.

The remaining difference between simulation and data on the tag side of the charmonium analysis is mitigated by using higher-order corrections at the generator level. The leading order \PYTHIA simulation is reweighted so that the \PB meson's \pt and \absy spectra match fixed-order plus next-to-leading logarithmic (FONLL) accuracy calculations, which are known to reproduce these spectra in data with high accuracy~\cite{Cacciari_1998}.

For the probe side, both in the open-charm and charmonium analyses, since the FONLL calculations cannot properly capture potential correlations between the probe-side and the trigger muon, simulated distributions of \pt and \absy are reweighted to match the measured ones at the reconstructed level. The weights are extracted for each \Bph, \Bzhs, and \Bsh decay by comparing simulation and background-subtracted data distributions, obtained using the $\mathit{_sPlot}$ technique~\cite{Pivk:2004ty}, and are parameterized via a second-order polynomial, as functions of \pt and $\absy$. Table~\ref{tab:mc_correction} summarizes the corrections applied to the simulated samples for the open-charm and charmonium channels. 

\begin{table}[htb]
\centering
\topcaption{Summary of MC corrections for the open-charm and charmonium channels.}
\begin{scotch}{ll}
Channel & MC correction \\
\hline
Open-charm & Reweight to data \\
[0.5em]
Charmonium (probe)  & Reweight to data \\
[0.5em]
\multirow{2}{*}{Charmonium (tag)}  & Muon trigger SF \\ 
& Reweighting to FONLL \\
\end{scotch}
\label{tab:mc_correction}
\end{table}

\section{Systematic uncertainties}
\label{sec:systematics}
In this analysis, systematic uncertainties are split into two categories. The first category is a bin-to-bin-uncorrelated systematic uncertainty, which depends on the \PB meson's \pt and $\absy$, and affects the shape of the measured PFR or \R values. The second type is bin-to-bin-correlated uncertainty, or global uncertainty, which does not affect the shape but impacts the central values of the measured PFRs. In the charmonium analysis, the uncertainties that affect only the global scale of the PFRs are not included in the measurement of \Rs and \Rsd, since this analysis only measures the shape versus \pt and \absy. Both types of uncertainties are included in the open-charm analysis and in the translation of the shape-only measurement of the PFRs in the charmonium analysis to an absolute normalization.

\subsection{Open-charm analysis}
The primary sources of systematic uncertainties and their effects on the PFR measurement in the open-charm analysis are summarized in Table~\ref{table:syst_opencharm}. Sources of systematic uncertainties correlated bin-to-bin and global uncertainties affecting only the central values of PFR are listed below. In this analysis, the systematic uncertainty related to the triggers is negligible, as their impact is found to cancel out in the PFRs for the probe-side analysis.

\begin{table*}[htb]
\centering
\topcaption{Sources and values of the systematic uncertainties affecting the measured PFRs in the open-charm analysis. The bin-to-bin-uncorrelated uncertainties are presented as ranges, which indicate the range of uncertainties across different \pt and \absy bins. The $\NA$ entries indicate that the uncertainty does not apply. The reported values are the relative systematic uncertainties in percent.}
\begin{scotch}{l c c c c}
 Source & $\fsfu$ ($\%$) & $\fsfd$ ($\%$)& $\fdfu ($\%$)$\\
\hline
$\Phi_{\text{PS}}, \,\abs{\Vus}\fK/(\abs{\Vud}\fpi), \, \tBz\!/\!\tBzs, \, \Na, \, \NF$ & 4.2 & 4.2 & \NA \\
Branching fractions & 5.4 & 4.0 & 4.2\\ 
Tracking efficiency & 2.1 & \NA & 2.1 \\
BDT performance & 1.2 & 2.2 & 2.2 \\
$\PBzs$ nonresonant subtraction & 0.9 & 0.9 & \NA \\
$r^{\pm,0}$ & \NA & \NA & 2.2 \\
\hline
Total global systematic uncertainty & 7.4 & 6.3 & 5.6 \\
\hline
Statistical uncertainty in simulation & 2.2--3.7& 2.2--3.7 & 2.1--3.6\\
Signal and background shapes  & 0.8--2.4 & 0.8--2.8& 0.7--2.8\\
Reweighting in \pt and \absy & 0.0--1.4& 0.0--1.6& 0.1--1.1\\
\hline
Total systematic uncertainty & $<$7.9 & $<$7.7 & $<$7.2\\
\end{scotch}
\label{table:syst_opencharm}
\end{table*}

The following global uncertainties are considered:
\begin{itemize}
\itemsep 0pt
\item Uncertainties in the external inputs to the PFR equations presented in Table~\ref{tab:inputsh}.

\item The uncertainty associated with the track reconstruction efficiency. Because \PBzs and \PBz meson decays have one more track than the \PBp decay, an uncertainty from a single-track reconstruction efficiency is assigned to the efficiency ratios, $\epsilon_{\PBp}/\epsilon_{\PBzs}$ and $\epsilon_{\PBz}/\epsilon_{\PBp}$. This uncertainty of 2.1$\%$~\cite{CMS-DP-2022-012} is treated as independent of \pt and \absy in this analysis, and therefore is considered a global uncertainty.

\item The uncertainty associated with the BDT selection procedure. Differences in the BDT efficiencies between simulation and data in each open-charm decay channel range from 1.2\% to 2.2\%. Since this uncertainty is small compared to the statistical and other systematic uncertainties, we assign the difference as the corresponding uncertainty but do not apply a correction. This uncertainty is found to be independent of \pt and \absy.

\item The uncertainty in the $\PBzs$ background subtraction to account for nonresonant $\PKp\PKm$ contribution to the \PDms decay. The expected amount of the nonresonant contribution within the $\pm10 \MeV$ mass window around the nominal $\PGf$ meson mass is found to be $5.6\pm0.9\%$~\cite{BaBar:2010wqe}, independent of \pt and \absy.

\item The measurement of \fdfu includes a 2.2\% uncertainty from the evaluation of $r^{\pm,0} = \BR(\PGUP{4S}\!\to\! \PBp\PBm)/\BR(\PGUP{4S}\!\to\! \PBz\PABz) = 1.057 \pm 0.023$~\cite{Bernlochner:2023bad}.
\end{itemize}
The overall bin-to-bin-correlated systematic uncertainty, calculated as the sum in quadrature of the individual components, ranges from
5.6\% to 7.4\%,
depending on the PFR\@.

For bin-to-bin-uncorrelated systematic uncertainties, the following sources are considered:
\begin{itemize}
\itemsep 0pt
\item The uncertainty in the signal efficiency arising from the limited number of simulated events is 1.7--3.7\% depending on the bin and the channel.

\item The uncertainty in the signal yields due to the parameterization of signal and background shapes is estimated by varying the functional forms presented in Table~\ref{tab:fit} to their alternative shapes and independently comparing the results obtained with the nominal function to those from the alternatives. A Johnson function is considered as an alternative shape for the signal, a Student-$t$ distribution~\cite{student-t} for the Cabibbo-suppressed background, a hyperbolic tangent function for the partially reconstructed background, and a first-order Chebyshev polynomial for the combinatorial backgrounds. The corresponding uncertainties vary between 0.7\% and 2.8\%, depending on the channel and the bin.

\item The uncertainty in the signal efficiency associated with the correction factors applied to simulated events. For the probe-side analysis, the \pt and \absy distributions from simulation are reweighted to match the measured distributions. The uncertainty is estimated by comparing the efficiency ratios $\epsilon_{\PBp}/\epsilon_{\PBzs}$, $\epsilon_{\PBz}/\epsilon_{\PBzs}$, and $\epsilon_{\PBz}/\epsilon_{\PBp}$ before and after applying the weights in \pt and \absy bins, and is less than 1.6\% for all processes and bins.
\end{itemize}
The overall bin-to-bin uncorrelated systematic uncertainties range from 2.0--4.7\%, depending on the channel and the bin.

\subsection{Charmonium analysis}
For the charmonium analysis, the dominant bin-to-bin uncorrelated systematic uncertainties in the tag-side and probe-side categories are listed in Table~\ref{table:syst_charm}. Several potential sources of systematic uncertainty, including those related to the muon reconstruction and selection efficiencies, are neglected in this analysis because they are found to cancel out in the efficiency ratios. For the probe-side category, as in the open-charm analysis, systematic uncertainty related to the triggers is not included, as its impact on the ratios is negligible.

\begin{table*}[htb]
\centering
\topcaption{Sources of bin-to-bin uncorrelated systematic uncertainty affecting the measured \Rs, \Rsd, and \fdfu values for the tag-side and probe-side categories in the charmonium analysis. The uncertainties are presented as ranges, which indicate the range of the uncertainties across different \pt and \absy bins. The $\NA$ entries indicate that the uncertainty does not apply. The reported values are the relative systematic uncertainty in percent.}
\begin{scotch}{l c c c c c c c}
\multirow{2}{*}{Source} & \multicolumn{3}{c}{Tag} & \quad &\multicolumn{3}{c}{Probe} \\
 & $\Rs$ ($\%$) & $\Rsd$ ($\%$)& \fdfu ($\%$) & & $\Rs$ ($\%$) & $\Rsd$ ($\%$) & \fdfu ($\%$)\\
\hline
Stat. uncert. in simulation  & 1.1--5.1 & 1.3--7.1 & 1.1--5.6 & & 2.7--5.7  & 2.6--5.5 & 2.3--4.9\\
Signal and bkg. shapes       & 0.8--4.4 & 0.6--4.7 & 0.5--4.2 & & 1.3--10.1 & 0.9--9.4 & 0.9--6.6\\
Trigger SF                   & 1.9--2.6 & 1.9--2.7 & 1.9--2.7 & & \NA & \NA & \NA \\
Reweighting in \pt and \absy & 0.2--1.6 & 0.1--1.2 & 0.2--1.9 & & 0.1--1.8  & 0.3--2.2 & 0.2--1.7\\
\hline
Total syst. uncertainty      & $<$7.5   & $<$7.8   & $<$ 7.7  & & $<$11.7   & $<$11.1  & $<$8.3\\
\end{scotch}
\label{table:syst_charm}
\end{table*}

The following sources of systematic uncertainty in the \Rs, \Rsd and \fdfu values are considered:
\begin{itemize}
\itemsep 0pt
\item The uncertainty in the signal efficiency from the limited number of simulated events.

\item The uncertainty in the signal yields due to the parameterization of signal and background shapes. As in the open-charm analysis, the uncertainty is estimated by varying the functional forms presented in Table~\ref{tab:fit} to alternative shapes. For the charmonium analysis, a triple-Gaussian function with a common mean is considered as an alternative shape for the signal and a first-order Chebyshev polynomial for the combinatorial backgrounds. Uncertainties associated with the Cabibbo-suppressed and partially reconstructed backgrounds are found to be negligible and therefore not included.

\item The uncertainty in the signal efficiency associated with the trigger SF for the tag-side category. The uncertainty is computed by varying the SF up and down by its uncertainty and determining the impact on the signal yields.

\item The uncertainty in the signal efficiency associated with the correction factors applied to the simulated events. For the tag-side events, simulated \pt and $\absy$ distributions are reweighted based on the FONLL calculations, while for the probe-side they are reweighted to match the measured distributions. In both cases, the uncertainty is computed by comparing the efficiency ratios $\epsilon_{\PBp}/\epsilon_{\PBzs}$ and $\epsilon_{\PBz}/\epsilon_{\PBzs}$ before and after applying the weights.

\end{itemize}
The overall uncertainties vary from about 2\% to 10\%, depending on the channel and whether the events are from the tag-side or probe-side. Significant uncertainties on the tag side are mainly associated with the low-\pt bins, where a fine binning of 1\GeV is used. On the probe side, large uncertainties are primarily associated with the high-rapidity region. The charmonium analysis employs a finer binning in \absy, splitting into 9 bins, compared to the 6 bins in the open-charm analysis.

For the measurement of \fdfu, the following global uncertainties are considered:
\begin{itemize}
\itemsep 0pt
\item An uncertainty of 2.1\% associated with the track reconstruction efficiency~\cite{CMS-DP-2022-012}. Because \PBz decays have one more track than the \PBp process, the 2.1\% uncertainty is assigned to the efficiency ratio, $\epsilon_{\PBz}/\epsilon_{\PBp}$. This uncertainty is treated as a global uncertainty.

\item An uncertainty of 4.3\% associated with the branching fraction ratio between the \Bpcs and \Bzcs channels.

\item An uncertainty of 2.2\% from $r^{\pm,0} = 1.057 \pm 0.023$~\cite{Bernlochner:2023bad}.
\end{itemize}

\section{Results}
\label{sec:results}
In this section, we describe the following main measurements performed in this analysis:
\begin{itemize}
\itemsep 0pt
\item Measurement of the \PB meson PFRs (\fsfu and $\fsfd$), using the open-charm decays \Bphs, \Bzhs, and \Bshs.

\item Measurement of the relative PFR ($\Rs$ and \Rsd) using the charmonium decays \Bpcs, \Bzcs, and \Bzcs.

\item Measurement of \fdfu in both open-charm and charmonium channels without the assumption of isospin invariance.

\item Measurement of the normalization factors needed for translation of the \Rsd and \Rs values to the corresponding PFRs.
\end{itemize}

In the open-charm decay channels, the PFRs \fsfu and \fsfd are directly measured using the theoretical calculations of the ratio of the open-charm branching fractions for the \PBz and \PBzs mesons, as discussed in Section~\ref{sec:strategy}. For the charmonium decay channels, the relative production fraction ratios \Rsd and \Rs are measured to determine the trend of the PFRs in \pt or \absy with an improved statistical precision compared to previous analyses. The value of \fdfu is also directly determined both in the open-charm and charmonium channels to probe isospin invariance in \PB meson production.

Finally, the results obtained from the probe-side category in the charmonium channels are combined with those derived from the open-charm channels. This combination allows for the estimation of the absolute normalization of the \Rsd and \Rs values relative to \fsfd and \fsfu. The normalization is derived using precise measurements of the ratios of branching fractions from charmonium to open-charm decays, which improve upon several world-average values~\cite{ParticleDataGroup:2024cfk}. With this absolute normalization established on the probe side, the PFRs are extracted from the charmonium decays on the tag side. The details of which channels are used for each measurement are summarized in Table~\ref{tab:measurements}.

In addition to measuring the PFRs versus \pt and \absy, we also determine average values, which come from averaging the values in bins of \pt, using the inverse of the squared uncertainties as weights.  The statistical uncertainties for each average are obtained by calculating the weighted average using the statistical uncertainties only.  The central values of the averages come from the weighted average in which the weights come from summing the statistical and uncorrelated bin-by-bin systematic uncertainties in quadrature.  The global uncertainty is added, in quadrature, to the uncertainty of this result to obtain the total uncertainty.  The systematic uncertainty is obtained by subtracting, in quadrature, the statistical uncertainty from the total uncertainty.  Angle brackets are used to indicate an average in a \pt range where a \pt dependence is expected.

\begin{table}[htbp]
\centering
\topcaption{Summary of the channels employed in each measurement of \fsfu, \fsfd, \Rs, \Rsd, \fdfu, and in determining the absolute normalizations for \Rs and \Rsd. The $\NA$ entries indicate the channel is not included.}
\begin{scotch}{lccc}
\multirow{2}{*}{Measurements} & \multirow{2}{*}{Open-charm} &  \multicolumn{2}{c}{Charmonium} \\
  & & Probe & Tag \\
\hline
\(\fsfu, \fsfd\)                & Included & \NA       & \NA       \\
\(\Rs, \Rsd\)                   & \NA       & Included & Included \\
\(\fdfu\)                     & Included & Included & Included \\
\Rs, \Rsd norm. & Included & Included & \NA       \\
\end{scotch}
\label{tab:measurements}
\end{table}

\subsection{Measurement of \texorpdfstring{\PB}{B} meson PFRs using open-charm decays}
\begin{figure*}[bhtp]
\centering
\includegraphics[width=0.48\textwidth]{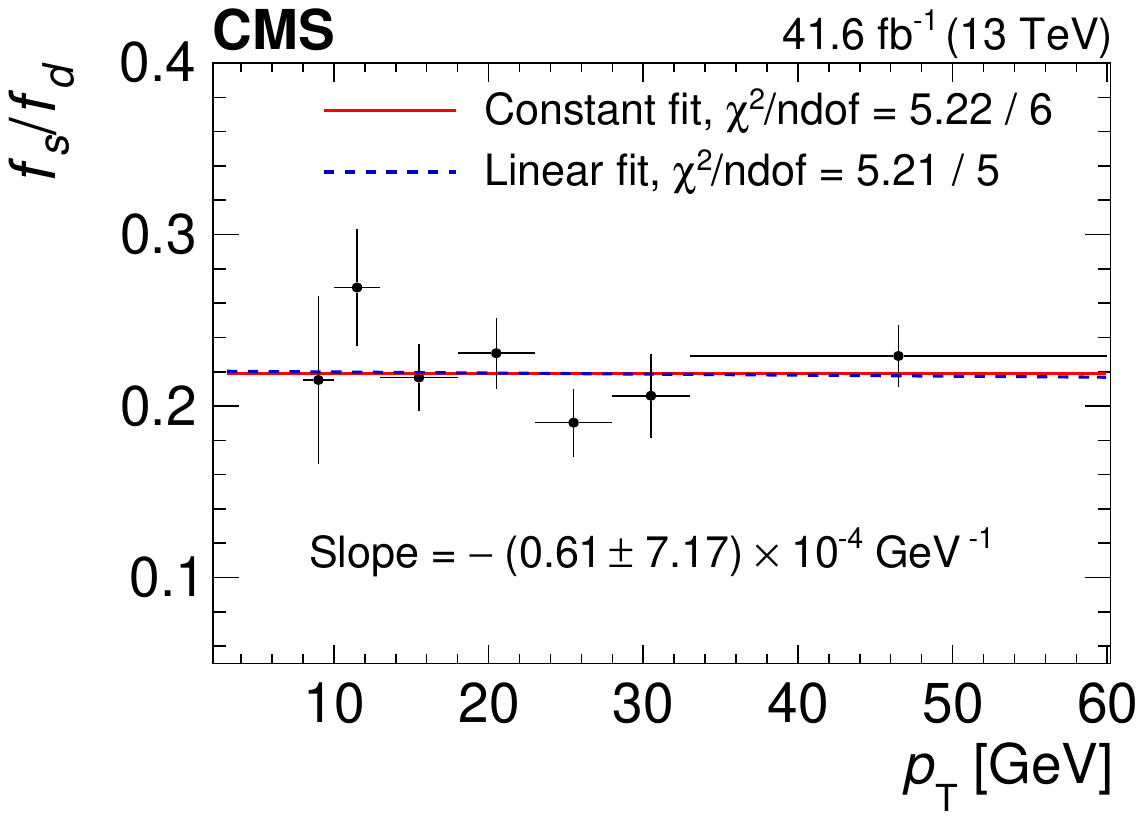}
\includegraphics[width=0.48\textwidth]{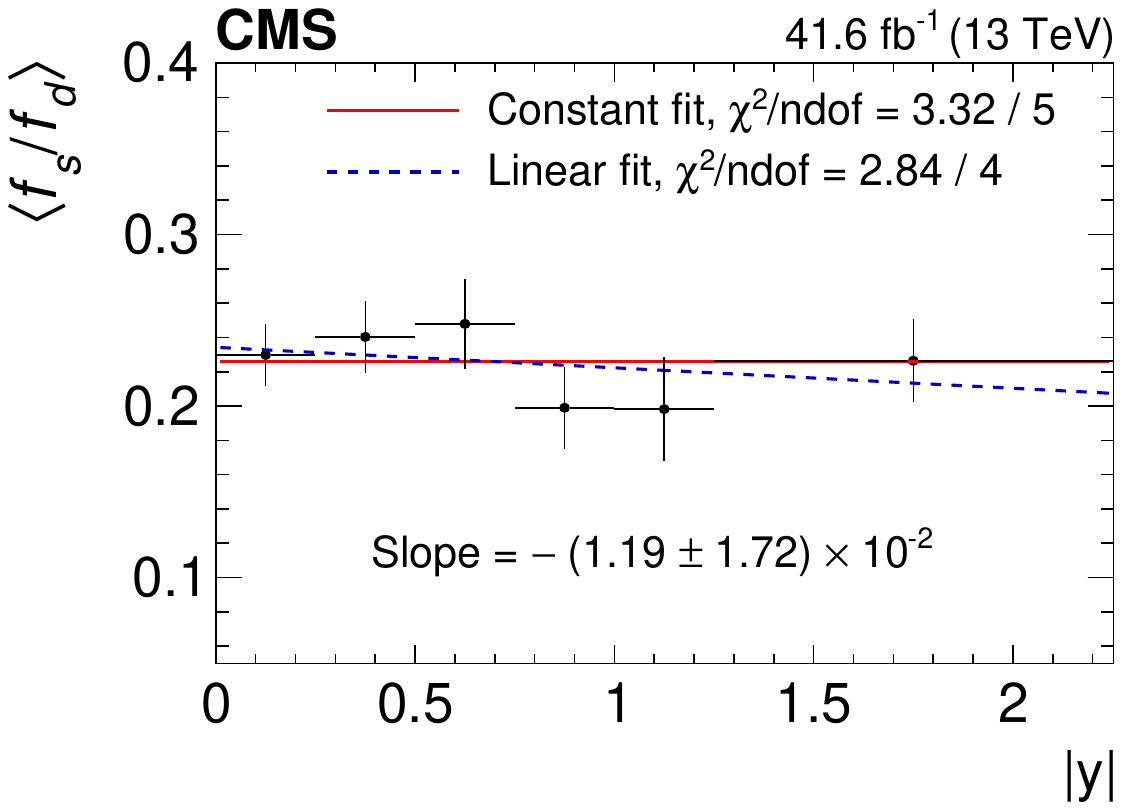}
\includegraphics[width=0.48\textwidth]{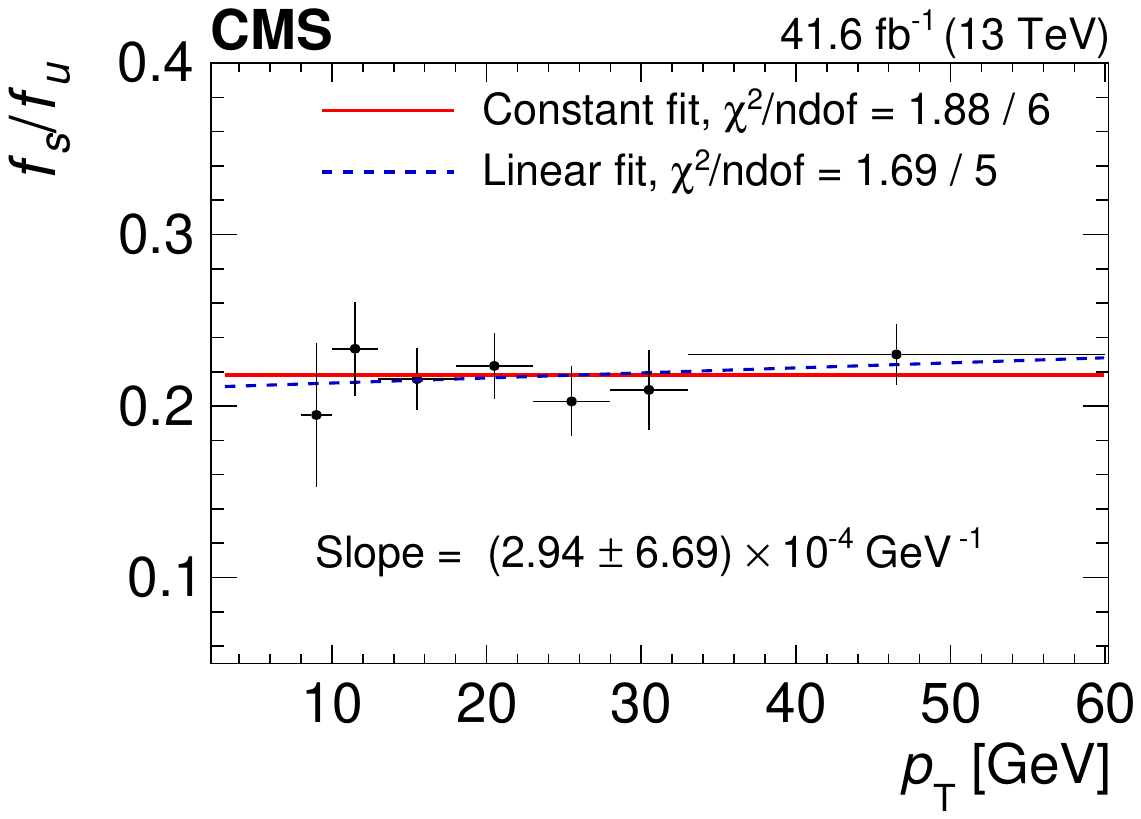}
\includegraphics[width=0.48\textwidth]{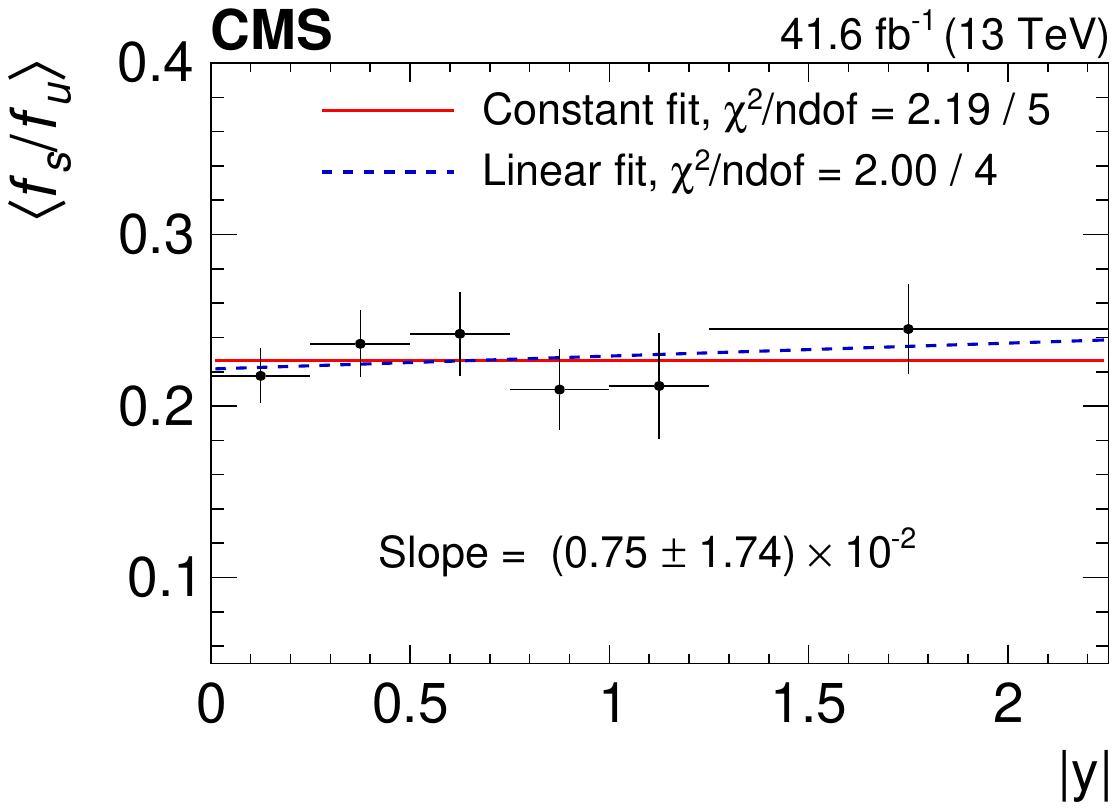}
\caption{The production fraction ratios  \fsfd (upper row) and \fsfu (lower row) as functions of $\pt$ (left column) and $\absy$ (right column), measured using open-charm decays. The vertical error bars show the combined statistical and bin-to-bin uncorrelated systematic uncertainty. The global uncertainties are not included. The red line is the average over the full reconstructed kinematic range, and the blue dashed line is the linear fit result. The $\chi^2$ per degree of freedom ($\chi^2/\mathrm{ndof}$) values for the constant fit and linear fit are given. Parameters inside angle brackets are averages over \pt where the absolute normalization may be affected by the \pt dependence.}
\label{fig:PFR}
\end{figure*}

The measured PFRs, \fsfu and \fsfd, are shown as functions of \pt and \absy in Fig.~\ref{fig:PFR}. For each measured PFR, hypotheses regarding both a linear dependence and no dependence on the \PB meson kinematics are tested. The results indicate that there is no evidence of a linear dependence of the PFRs on either \pt or \absy within the range of this measurement.

The averages over the full reconstructed kinematic range of $8<\pt<60\GeV$ and $\abs{y}<2.25$ from the PFR \pt values are:
\begin{align*}
\langle\fsfd\rangle & = 0.219 \pm 0.008\stat \pm 0.014\syst, \\
\langle\fsfu\rangle & = 0.218 \pm 0.008\stat \pm 0.015\syst.
\end{align*}
Due to the previously observed low-\pt trend in the PFRs~\cite{LHCb:2021qbv,CMS:2022wkk}, any values of \fsfd or \fsfu obtained from integrating over the full \pt range are considered to be {\it effective\/} values. However, the trend of the PFRs in the \absy distribution remains unaffected, as the \absy distributions are not correlated with the \pt distributions within the kinematic range of this study.

The recent CMS study of \PB meson PFRs in charmonium decays~\cite{CMS:2022wkk} showed that there exists a \pt dependence of \fsfu and \fsfd at low \pt, followed by a plateau for $\pt \gtrsim 18\GeV$.  The present measurement does not have sufficient precision to probe the low-\pt dependence because of the lower event count in the relevant \pt bins compared to the charmonium decay mode of the previous CMS study, which is based on a larger data set collected with regular triggers. However, in the high-\pt region, where the PFRs are observed to be independent of \PB meson kinematics, they can be measured with high precision. The average over the $\pt>18 \GeV$ region yields:
\begin{align*}
\fsfd & = 0.215 \pm 0.010\stat \pm 0.014\syst,   \\
\fsfu & = 0.218 \pm 0.009\stat \pm 0.016 \syst. 
\end{align*}
The measured high-\pt value of \fsfd is within one standard deviation of the corresponding LEP measurement of $\fsfd = 0.249 \pm 0.023$~\cite{HFLAV:2019otj} at high $\pt$ (around 40 \GeV).

\subsection{Measurement of \texorpdfstring{$\Rs$ and $\Rsd$}{Rs and Rsd} using charmonium decays}
\begin{figure*}[bhtp]
\centering
\includegraphics[width=0.48\textwidth]{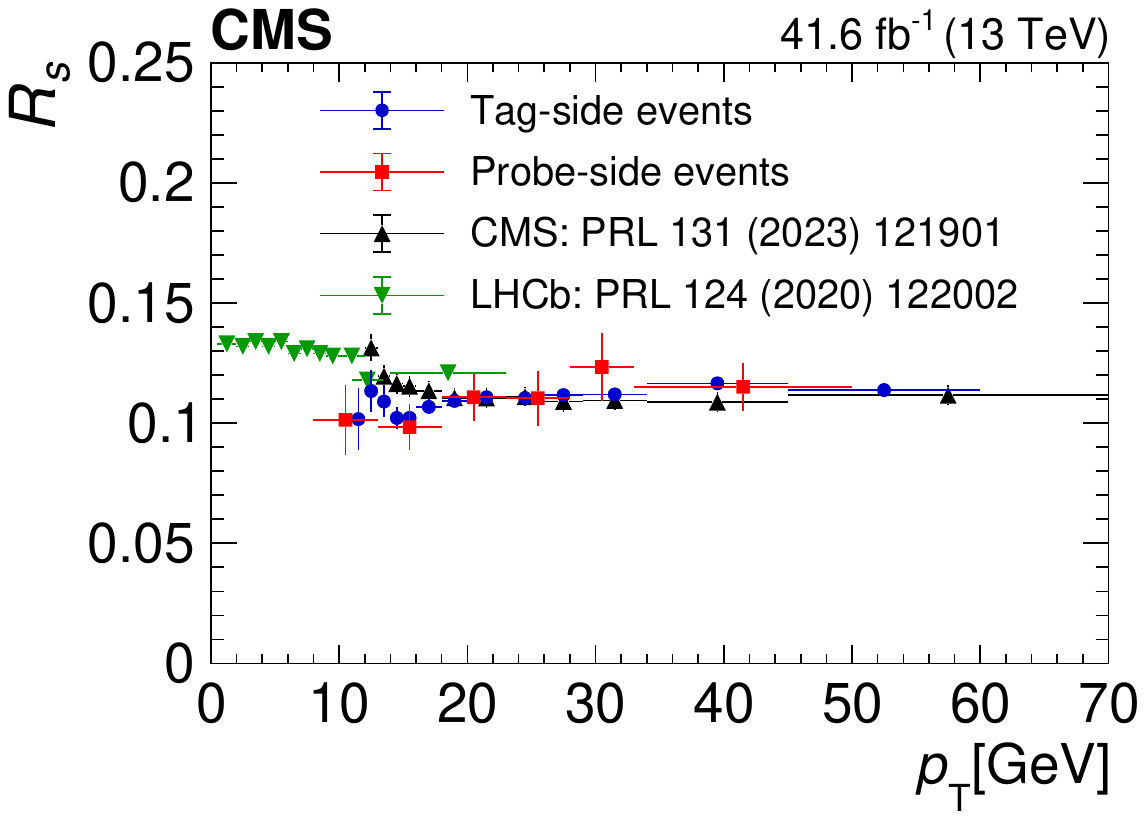}
\includegraphics[width=0.48\textwidth]{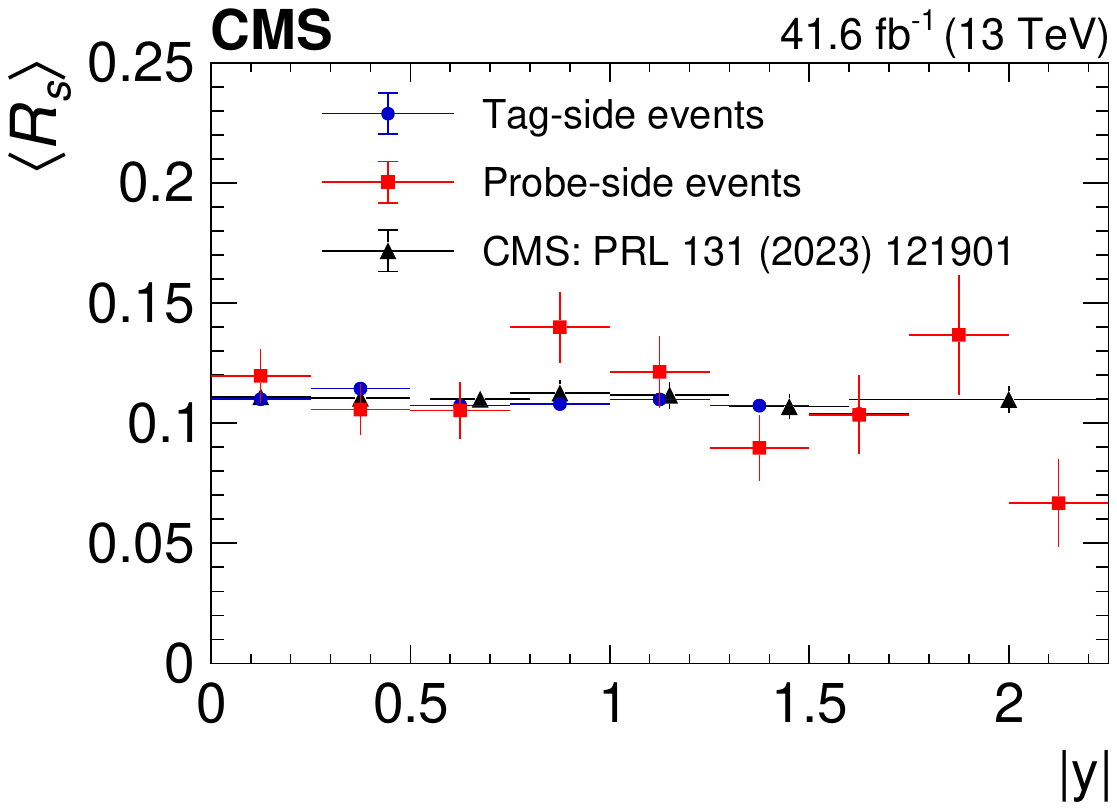}
\includegraphics[width=0.48\textwidth]{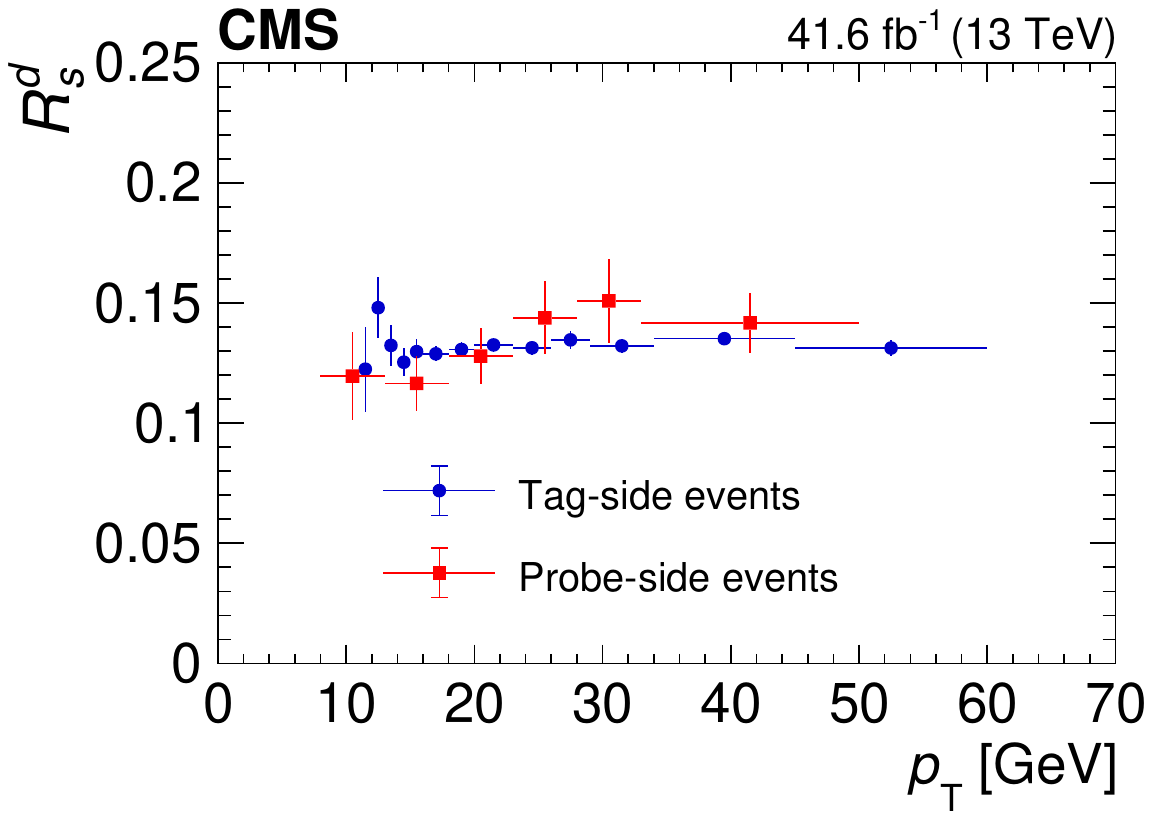}
\includegraphics[width=0.48\textwidth]{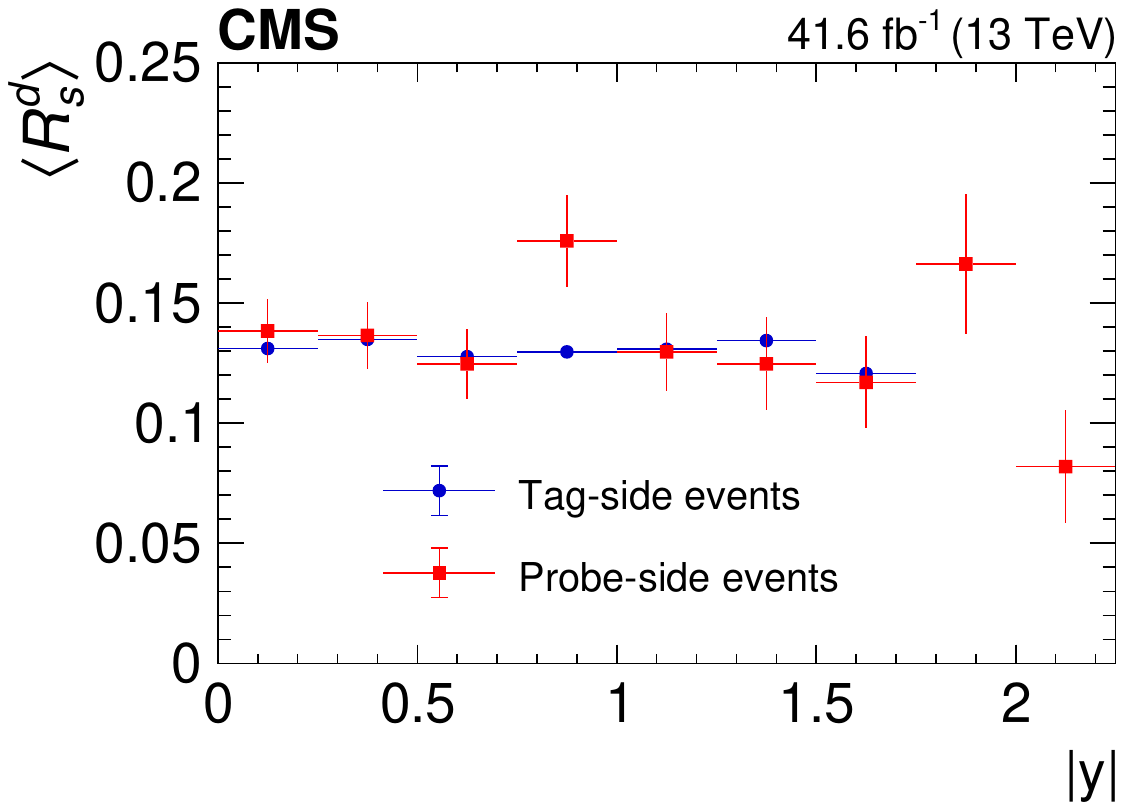}
\caption{The relative production fraction ratios $\Rs$ (upper row) and $\Rsd$ (lower row) as functions of \pt (left column) and \absy (right column), measured using charmonium decays. For comparison, the recent CMS~\cite{CMS:2022wkk} and LHCb~\cite{LHCb:2019lsv} measurements are also presented. The error bars include both statistical and bin-to-bin uncorrelated systematic uncertainties. Values enclosed in angle brackets represent averages over \pt, where the absolute normalization may be affected by the \pt dependence.}
\label{fig:R_values}
\end{figure*}

The results of the measurements of \Rs and \Rsd are presented in bins of \pt and \absy in Fig. \ref{fig:R_values} for both tag- and probe-side categories. While the measured \Rs and \Rsd values do not exhibit any statistically significant dependence on either \pt or \absy, they are consistent within the uncertainties with the \pt-dependence at low \pt, followed by a flat high-\pt trend in both \Rs and \Rsd observed in earlier analyses~\cite{LHCb:2019lsv,CMS:2022wkk}.

\subsection{Measurement of \texorpdfstring{$\fdfu$}{fdfu} using both the open-charm and charmonium decays}
\begin{figure}[bhtp]
\centering
\includegraphics[width=0.48\textwidth]{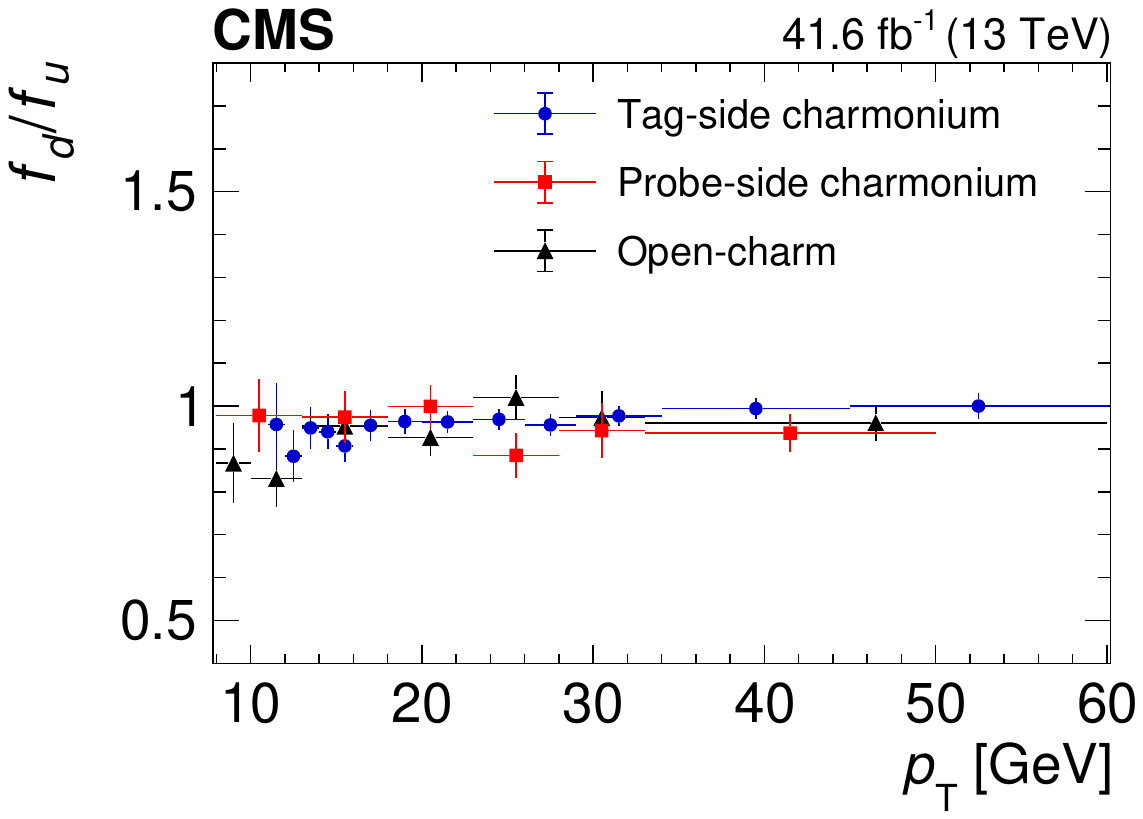}
\includegraphics[width=0.48\textwidth]{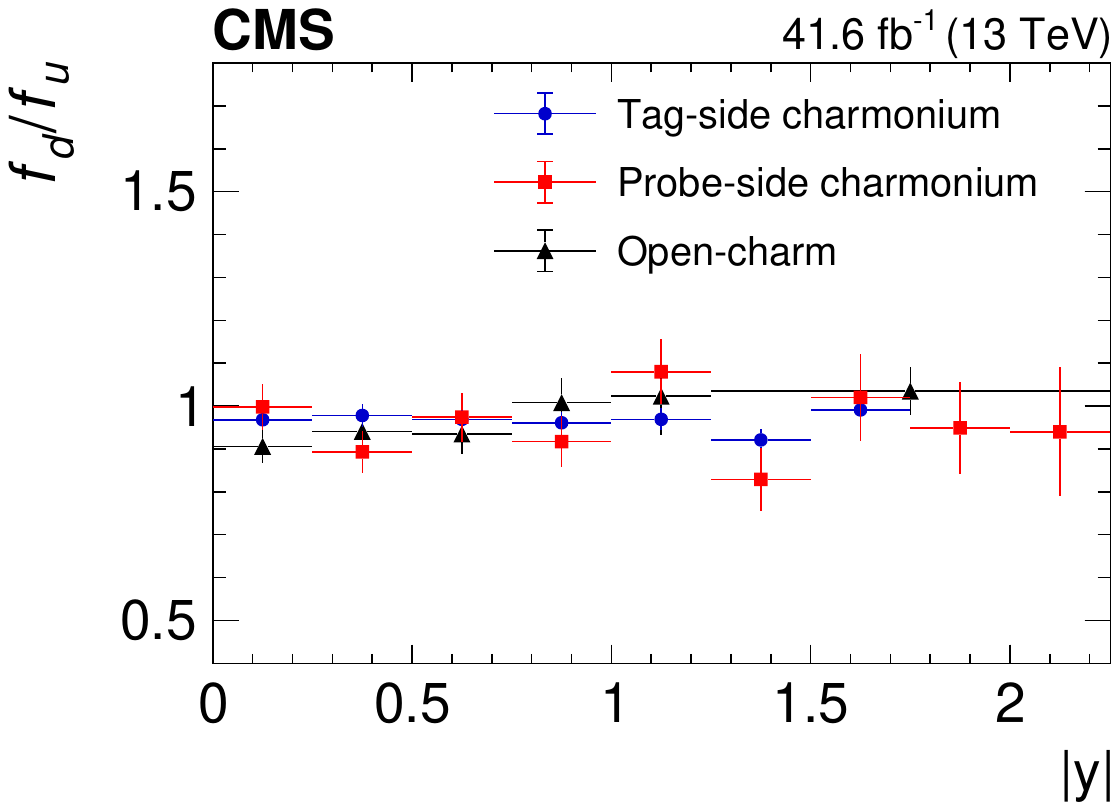}
\caption{The production fraction ratios $\fdfu$ as functions of \pt (\cmsLeft) and \absy (\cmsRight) measured using both the open-charm and charmonium decays, without an assumption of isospin invariance. The error bars include both statistical and bin-to-bin uncorrelated systematic uncertainties. The global uncertainties of 5.7\% for the open-charm analysis and 4.8\% for both charmonium analyses are not shown.}
\label{fig:fdfu_values}
\end{figure}

The measured values of \fdfu obtained from Eq.~(\ref{eq:fdfu}) for the open-charm analysis and Eq.~(\ref{eq:fdfuc}) for the charmonium analysis are presented as functions of \pt and \absy in Fig.~\ref{fig:fdfu_values}.  The results from each of the three analyses are consistent with each other and show no dependence on \pt or \absy.  The results are also consistent with a previous CMS analysis~\cite{CMS:2022wkk}, which covered the range of $12<\pt<70$\GeV and $\absy<2.4$.  The average values are calculated to be:
\begin{align*}
\fdfu(\text{open-charm}) & = 0.946 \pm 0.016 \pm 0.055, \\
\fdfu(\text{charmonium, tag}) & = 0.965 \pm 0.002 \pm 0.047,\\
\fdfu(\text{charmonium, probe}) & = 0.949 \pm 0.019 \pm 0.047,
\end{align*}
where the first and second uncertainties are statistical and systematic, respectively.

\begin{figure}[bhtp]
\centering
\includegraphics[width=\figwid]{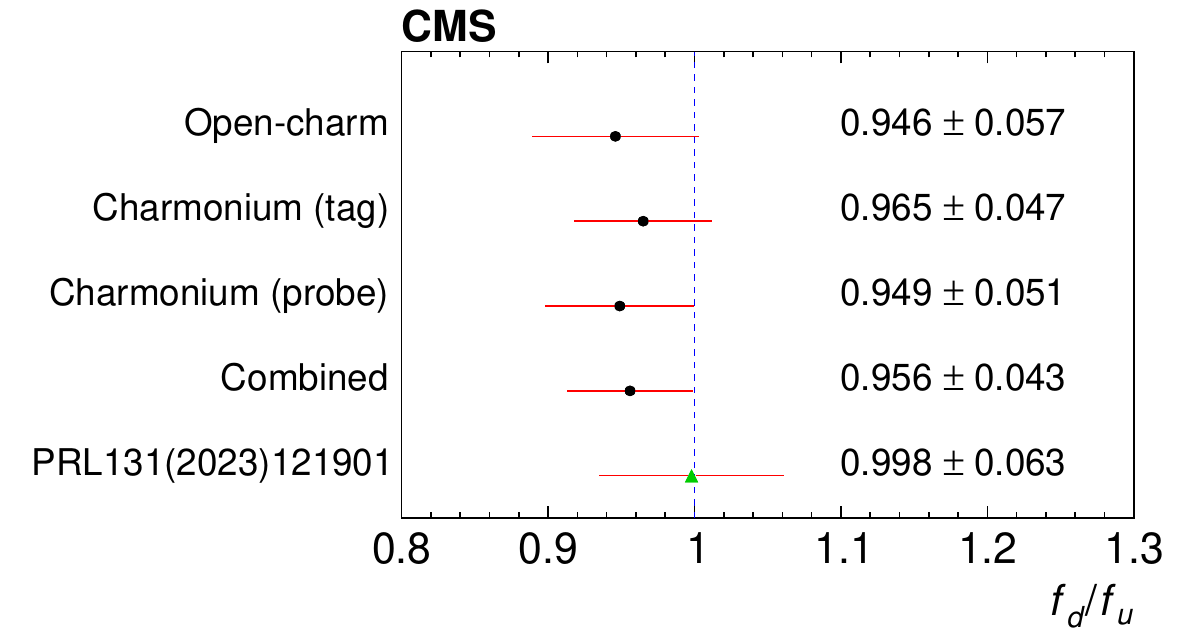}
\caption{Comparison of the $\fdfu$ measurements across different channels. The ``Combined'' value, shown above the black dashed line, is obtained from a combined $\chi^{2}$ fit of the measurements from two statistically independent channels: the open-charm and tag-side charmonium. Additionally, the plot presents the previous CMS result~\cite{CMS:2022wkk} for comparison. The blue dashed line at unity corresponds to $f_{\PQd} = f_{\PQu}$.}
\label{fig:fdfu_comparison}
\end{figure}

We calculate a weighted average of the open-charm and charmonium \fdfu values, accounting for correlations within each analysis, including systematic uncertainties from external inputs and the BDT, as well as correlations between the two analyses, such as the uncertainties from the tracking efficiency and $r^{\pm,0}$. The combined \fdfu value is:
\begin{equation*}
\fdfu = 0.956 \pm 0.043.
\end{equation*}
The \fdfu value is compatible with unity within a 5\% precision, consistent with isospin invariance in $\PB$ meson production at hadron colliders. Figure~\ref{fig:fdfu_comparison} summarizes the CMS \fdfu measurements across decay channels, with the uncertainties including significant contributions from common systematic uncertainties.

We also measure $\fdfu$ for particles and antiparticles separately.  Combining the results from the open-charm and both charmonium analyses, we obtain:
\begin{align*}
f_\PBp/f_\PBz & = 0.947 \pm 0.044, \\
f_\PBm/f_\PABz & = 0.957 \pm 0.044.
\end{align*}
Taking the ratio of the two ratios returns $0.989 \pm 0.015$, where the uncertainty does not include any global systematic uncertainties as they cancel out in this double ratio.

\subsection{Measurement of \texorpdfstring{\PB}{B} branching fractions}
The ratio of the efficiency-corrected yields between the charmonium and open-charm decays corresponds to the ratio of the branching fractions of the charmonium to open-charm decays. We use the probe-side charmonium sample to minimize bias from the triggers and match the \pt binning of the open-charm sample to that of the probe-side charmonium sample to remove the effect of any \pt dependence.  The weighted averages of the ratios of the corrected yields in the \pt bins are used to extract the ratios of the branching fractions:
\ifthenelse{\boolean{cms@external}}
{
\begin{widetext}
\begin{align}
&\frac{\BR(\Bpcs)}{\BR(\Bphs)} = \frac{\Ncorr(\Bpc)}{\Ncorr(\Bph)} \frac{\BR(\Dzh)}{\BR(\JPsi\!\to\!\MM)} = 0.213 \pm 0.007, \label{eq:br_measured1} \\[4pt]
&\frac{\BR(\Bscs)}{\BR(\Bshs)} = \frac{\Ncorr(\Bsc)}{\Ncorr(\Bsh)} \frac{\BR(\Dsh)}{\BR(\JPsi\!\to\!\MM)} = 0.354 \pm 0.019, \label{eq:br_measured2} \\[4pt]
&\frac{\BR(\Bzcs)}{\BR(\Bzhs)} =  \frac{\Ncorr(\Bzc)}{\Ncorr(\Bzh)} \frac{\BR(\Dmh)}{\BR(\KsKpi)\BR(\JPsi\!\to\!\MM)} =  0.468 \pm 0.021, \label{eq:br_measured3}
\end{align}
\end{widetext}
}
{
\begin{equation}
\frac{\BR(\Bpcs)}{\BR(\Bphs)} 
= \frac{\Ncorr(\Bpc)}{\Ncorr(\Bph)} 
\frac{\BR(\Dzh)}{\BR(\JPsi\!\to\!\MM)} 
= 0.213 \pm 0.007,
\label{eq:br_measured1}
\end{equation}
\begin{equation}
\frac{\BR(\Bscs)}{\BR(\Bshs)} 
= \frac{\Ncorr(\Bsc)}{\Ncorr(\Bsh)} 
\frac{\BR(\Dsh)}{\BR(\JPsi\!\to\!\MM)}  
= 0.354 \pm 0.019,
\label{eq:br_measured2}
\end{equation}
\begin{equation}
\begin{split}
\frac{\BR(\Bzcs)}{\BR(\Bzhs)} &=  \frac{\Ncorr(\Bzc)}{\Ncorr(\Bzh)} \frac{\BR(\Dmh)}{\BR(\KsKpi)\BR(\JPsi\!\to\!\MM)}\\
& =  0.468 \pm 0.021,
\label{eq:br_measured3}
\end{split}
\end{equation}
}
where the results in Eqs.~(\ref{eq:br_measured2}) and (\ref{eq:br_measured3}) have been corrected to remove the $S$-wave contributions in the mass ranges used to select the $\Pphi$ and $\PKstn$ decays, respectively.  The $S$-wave contributions are estimated as $(1.0 \pm 0.2)\%$ for Eq.~(\ref{eq:br_measured2})~\cite{LHCb:2021qbv} and $(2.2 \pm 0.2)\%$ for Eq.~(\ref{eq:br_measured3}), with the latter obtained from considering the $\PKstP{892}{}^0$ and $\PKstDzP{700}$ contributions in the model of Ref.~\cite{Belle:2014nuw}, including the uncertainty from varying these two fit fractions.  The uncertainties in Eqs.~(\ref{eq:br_measured1}--\ref{eq:br_measured3}) include both statistical and systematic components.  The global uncertainties included are those from the branching fractions appearing in Eqs.~(\ref{eq:br_measured1}--\ref{eq:br_measured3}), the BDT performance and $\PBzs$ nonresonant component subtraction (Table~\ref{table:syst_opencharm}), and a 2.1\% tracking efficiency uncertainty.  While each branching fraction ratio measurement consists of the same number of charged particles in the numerator and the denominator, two of the charged particles are muons for the charmonium channels while all of the charged particles are hadrons for the open-charm channels.  This leads to only a partial cancellation of the tracking efficiency uncertainty and the value of 2.1\% is selected as the midpoint between full cancellation and no cancellation.

Our measured branching fraction ratios are consistent with the world-average values of $\BR(\Bpcs)/\BR(\Bphs) = 0.221 \pm 0.006$, $\BR(\Bscs)/\BR(\Bshs) = 0.346 \pm 0.021$, and $\BR(\Bzcs)/\BR(\Bzhs) = 0.496 \pm 0.025$~\cite{ParticleDataGroup:2024cfk}, with similar precision.

\subsection{Measurement of the normalization for the efficiency-corrected yield ratios \texorpdfstring{\R}{R}}
The branching fractions measured in Eqs.~(\ref{eq:br_measured1},\ref{eq:br_measured2},\ref{eq:br_measured3}) can be used to extract the factors necessary to translate \Rs and \Rsd to the absolute PFRs, \fsfu and \fsfd.
These normalization factors, $c_{\PQs\PQd}$ and $c_{\PQs\PQu}$, are calculated as:
\ifthenelse{\boolean{cms@external}}
{
\begin{equation}
\begin{split}
c_{\PQs\PQd} & =  (\fsfd) \text{(open-charm)}/\Rsd \text{(charmonium)} \\
& = \Phi_{\text{PS}} \frac{\abs{\Vus}^2}{\abs{\Vud}^2} \frac{f^2_\PK}{f^2_\PGp} \frac{\tBz}{\tBzs}\frac{1}{\Na\NF} \\
& \qquad \times \frac{\BR(\Bzhs)}{\BR(\BzhKs)}  \frac{\BR(\Dmh)}{\BR(\Dsh)\BR(\phiKK)}\\
& \qquad \times \frac{\Ncorr(\Bsh)}{\Ncorr(\Bsc)}  \\
& \qquad \times \frac{\Ncorr(\Bzc)}{\Ncorr(\Bzh)} \\
& =  \Phi_{\text{PS}} \frac{\abs{\Vus}^2}{\abs{\Vud}^2} \frac{f^2_\PK}{f^2_\PGp} \frac{\tBz}{\tBzs}\frac{1}{\Na\NF} \\
& \qquad \times \frac{\BR(\Bzhs)}{\BR(\BzhKs)} \frac{\BR(\KsKpi)}{\BR(\phiKK)}\\
& \qquad \times \frac{\BR(\Bshs)}{\BR(\Bscs)}\frac{\BR(\Bzcs)}{\BR(\Bzhs)},
\label{eq:csd}
\end{split}
\end{equation}
\begin{equation}
\begin{split}
c_{\PQs\PQu} & =  (\fsfu)\text{(open-charm)}/\Rs\text{(charmonium)} \\
& = \Phi_{\text{PS}}  \frac{\abs{\Vus}^2}{\abs{\Vud}^2}  \frac{f^2_\PK}{f^2_\PGp} \frac{\tBz}{\tBzs}\frac{1}{\Na\NF} \\
& \qquad \times \frac{\BR(\Bphs)}{\BR(\BzhKs)} \frac{\BR(\Dzh)}{\BR(\Dsh)\BR(\phiKK)}\\
& \qquad \times \frac{\Ncorr(\Bsh)}{\Ncorr(\Bsc)} \\
& \qquad \times \frac{\Ncorr(\Bpc)}{\Ncorr(\Bph)} \\
& = \Phi_{\text{PS}}  \frac{\abs{\Vus}^2}{\abs{\Vud}^2} \frac{f^2_\PK}{f^2_\PGp} \frac{\tBz}{\tBzs}\frac{1}{\Na\NF} \\
& \qquad \times \frac{\BR(\Bphs)}{\BR(\BzhKs)} \frac{1}{\BR(\phiKK)} \\
& \qquad \times \frac{\BR(\Bshs)}{\BR(\Bscs)} \frac{\BR(\Bpcs)}{\BR(\Bphs)},
\label{eq:csu}
\end{split}
\end{equation}
}
{
\begin{equation}
\begin{split}    
c_{\PQs\PQd} & =  (\fsfd) \text{(open-charm)}/\Rsd \text{(charmonium)} \\
& = \Phi_{\text{PS}} \frac{\abs{\Vus}^2}{\abs{\Vud}^2} \frac{f^2_\PK}{f^2_\PGp} \frac{\tBz}{\tBzs}\frac{1}{\Na\NF} \frac{\BR(\Bzhs)}{\BR(\BzhKs)}  \frac{\BR(\Dmh)}{\BR(\Dsh)\BR(\phiKK)}\\
& \qquad \times \frac{\Ncorr(\Bsh)}{\Ncorr(\Bsc)}  \frac{\Ncorr(\Bzc)}{\Ncorr(\Bzh)} \\
&=  \Phi_{\text{PS}} \frac{\abs{\Vus}^2}{\abs{\Vud}^2} \frac{f^2_\PK}{f^2_\PGp} \frac{\tBz}{\tBzs}\frac{1}{\Na\NF} \frac{\BR(\Bzhs)}{\BR(\BzhKs)} \frac{\BR(\KsKpi)}{\BR(\phiKK)}\\
& \qquad \times \frac{\BR(\Bshs)}{\BR(\Bscs)} \frac{\BR(\Bzcs)}{\BR(\Bzhs)},
\label{eq:csd}
\end{split}
\end{equation}
\begin{equation}
\begin{split}
c_{\PQs\PQu} & =  (\fsfu)\text{(open-charm)}/\Rs\text{(charmonium)} \\
& = \Phi_{\text{PS}}  \frac{\abs{\Vus}^2}{\abs{\Vud}^2}  \frac{f^2_\PK}{f^2_\PGp} \frac{\tBz}{\tBzs}\frac{1}{\Na\NF} \frac{\BR(\Bphs)}{\BR(\BzhKs)} \frac{\BR(\Dzh)}{\BR(\Dsh)\BR(\phiKK)}\\
& \qquad \times \frac{\Ncorr(\Bsh)}{\Ncorr(\Bsc)}\frac{\Ncorr(\Bpc)}{\Ncorr(\Bph)} \\
& = \Phi_{\text{PS}}  \frac{\abs{\Vus}^2}{\abs{\Vud}^2} \frac{f^2_\PK}{f^2_\PGp} \frac{\tBz}{\tBzs}\frac{1}{\Na\NF} \frac{\BR(\Bphs)}{\BR(\BzhKs)} \frac{1}{\BR(\phiKK)}\\
& \qquad \times  \frac{\BR(\Bshs)}{\BR(\Bscs)} \frac{\BR(\Bpcs)}{\BR(\Bphs)},
\label{eq:csu}
\end{split}
\end{equation}
}
where the branching fractions from the last line of each equation are obtained from Eqs.~(\ref{eq:br_measured1},\ref{eq:br_measured2},\ref{eq:br_measured3}), while the branching fractions in the next-to-last line of each equation are from Ref.~\cite{ParticleDataGroup:2024cfk} and reproduced in Table~\ref{tab:inputsh}.
The results are:
\begin{align*}
c_{\PQs\PQd} & = 1.64 \pm 0.14, \\
c_{\PQs\PQu} & = 2.02 \pm 0.18.
\end{align*}
The uncertainties include both statistical and systematic components.  The global uncertainties included in the calculation of Eqs.~(\ref{eq:br_measured1},\ref{eq:br_measured2},\ref{eq:br_measured3}) are included except for those from $\BR(\JPsi\!\to\!\MM)$, $\BR(\phiKK)$, and the tracking efficiency, as these all cancel out.  Additional global uncertainties associated with all of the values on the next-to-last line of each equation are also included.

\begin{figure}[htbp]
\centering
\includegraphics[width=0.48\textwidth]{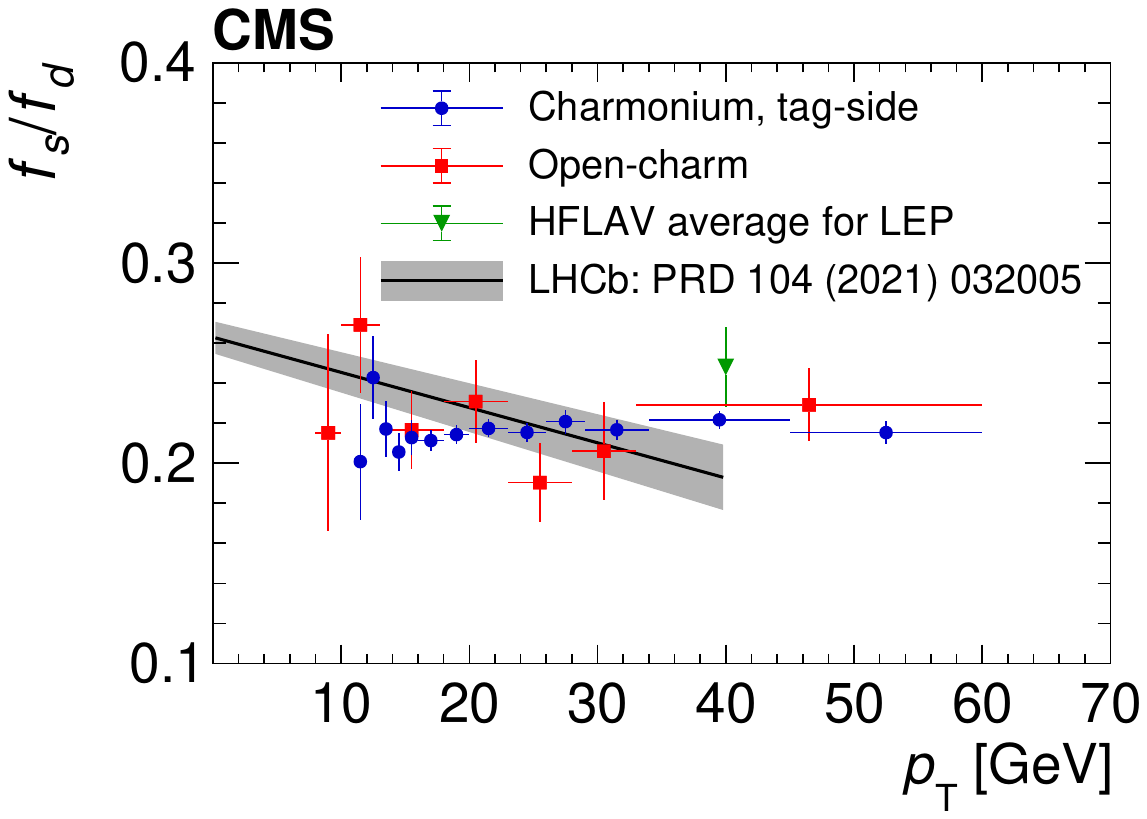}
\includegraphics[width=0.48\textwidth]{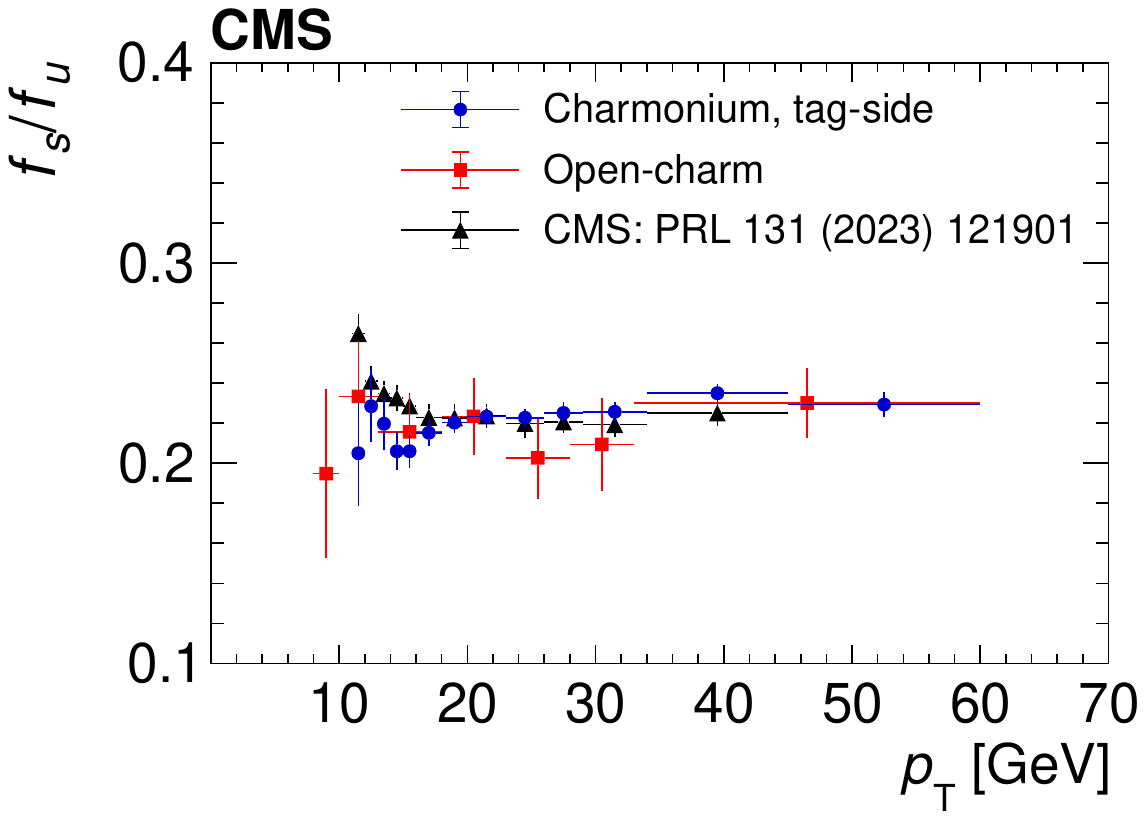}
\caption{The \Rsd and \Rs values measured in the charmonium analysis, converted to $\fsfd$ (\cmsLeft) and $\fsfu$ (\cmsRight) using the normalization factors $c_{\PQs\PQd}$ and $c_{\PQs\PQu}$, respectively. The error bars include both statistical and bin-to-bin uncorrelated systematic uncertainties. For comparison, PFR measurements from the open-charm analysis, LEP~\cite{HFLAV:2019otj}, and CMS~\cite{CMS:2022wkk} (converted to \fsfu using $c_{\PQs\PQu}$) are overlaid, along with the \pt trend of $\fsfd$ observed by LHCb~\cite{LHCb:2021qbv} with its uncertainties displayed as the gray band. The $\fsfd$ ($\fsfu$) global uncertainties are 6.3\% (7.4\%) for the open-charm channel and 8.4\% (8.9\%) for both the charmonium channel and the previous CMS result.}
\label{fig:Combined}
\end{figure}

\begin{figure}[bhtp]
\centering
\includegraphics[width=\figwid]{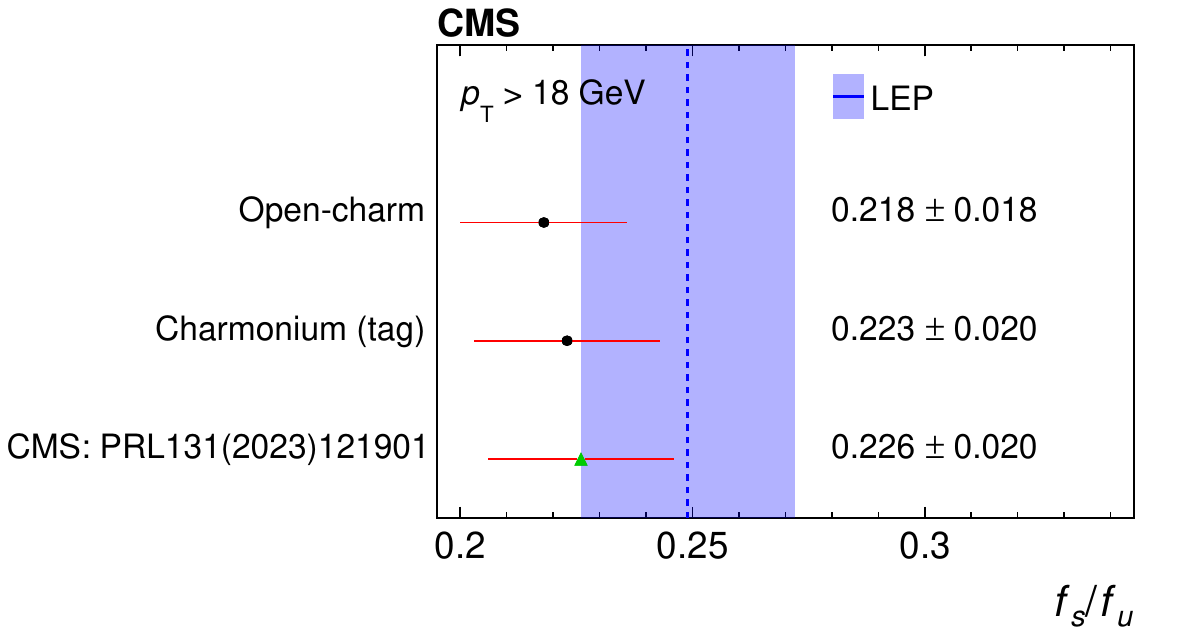}
\caption{Comparison of the $\fsfu$ measurements at high $\pt$ ($\pt > 18 \GeV$) across different channels. The blue dashed line and shaded band represent the LEP measurement and its uncertainty~\cite{HFLAV:2019otj}, respectively, included for comparison. The \Rs values obtained in the charmonium analysis in the tag-side category and the previous CMS measurement~\cite{CMS:2022wkk} are converted to $\fsfu$ using the normalization factor $c_{\PQs\PQu}$.}
\label{fig:fsfu_comparison}
\end{figure}

The \R values obtained in the tag-side charmonium analysis, as well as in an earlier CMS charmonium analysis~\cite{CMS:2022wkk}, are converted to \fsfd and \fsfu using these normalization factors and are compared in Fig.~\ref{fig:Combined} with the open-charm results, the LHCb~\cite{LHCb:2021qbv} linear fit function, and the LEP high-\pt value~\cite{HFLAV:2019otj}.  The measured PFR values \fsfd and \fsfu are consistent with previous measurements from LHCb, CMS, and LEP, including the observed \pt dependence at low \pt.  The converted PFR measurements from the charmonium analysis for the tag-side category over the full reconstructed kinematic range yield $\langle\fsfd\rangle = 0.216 \pm 0.018$ and $\langle\fsfu\rangle = 0.223 \pm 0.020$.  Restricting to the region of $\pt > 18$ GeV, where the PFRs are observed to be independent of \PB meson kinematics, gives $\fsfd = 0.217 \pm 0.019$ and $\fsfu = 0.226 \pm 0.020$. Figure~\ref{fig:fsfu_comparison} provides a summary of CMS \fsfu measurements in the high-\pt region, with the uncertainties dominated by systematic effects that are common to all measurements.

\section{Summary}
In summary, the relative production fractions of \PBp, \PBz, and \PBzs mesons have been measured in proton-proton collisions at $\sqrt{s} = 13\TeV$ using a special data set recorded in 2018 with the CMS experiment at the LHC, corresponding to an integrated luminosity of 41.6$\fbinv$. This data set, collected with high-rate triggers, provided an unprecedented unbiased sample of $10^{10}$ \PQb hadrons.

The production fraction ratios \fsfd, \fsfu, and \fdfu were measured as functions of \PB meson transverse momentum and absolute rapidity in the range of $8 < \pt < 60\GeV$ and $\absy < 2.25$ using the corrected yield of open-charm decays combined with known branching fractions and precise theoretical calculations.  These ratios were also measured in the charmonium decay channels, with the normalization of \fsfd and \fsfu in these channels obtained from the open-charm results.  These normalization factors were extracted as a double ratio of the decay modes, thereby canceling any potential \pt dependence of the production fraction ratios. This makes these quantities universal constants that can be applied to any charmonium analysis.

In addition to measuring the production fraction ratios as a function of the \PB meson kinematics, we also reported average values of these quantities, with the \fdfu result being a test of isospin invariance of \PB meson production.  The measurements were also used to extract \PBp, \PBzs, and \PBz branching fraction ratios of open-charm decays to charmonium decays, with precision similar to the current world-average values.

\begin{acknowledgments}
We congratulate our colleagues in the CERN accelerator departments for the excellent performance of the LHC and thank the technical and administrative staffs at CERN and at other CMS institutes for their contributions to the success of the CMS effort. In addition, we gratefully acknowledge the computing centers and personnel of the Worldwide LHC Computing Grid and other centers for delivering so effectively the computing infrastructure essential to our analyses. Finally, we acknowledge the enduring support for the construction and operation of the LHC, the CMS detector, and the supporting computing infrastructure provided by the following funding agencies: SC (Armenia), BMBWF and FWF (Austria); FNRS and FWO (Belgium); CNPq, CAPES, FAPERJ, FAPERGS, and FAPESP (Brazil); MES and BNSF (Bulgaria); CERN; CAS, MoST, and NSFC (China); MINCIENCIAS (Colombia); MSES and CSF (Croatia); RIF (Cyprus); SENESCYT (Ecuador); ERC PRG and PSG, TARISTU24-TK10 and MoER TK202 (Estonia); Academy of Finland, MEC, and HIP (Finland); CEA and CNRS/IN2P3 (France); SRNSF (Georgia); BMFTR, DFG, and HGF (Germany); GSRI (Greece); NKFIH (Hungary); DAE and DST (India); IPM (Iran); SFI (Ireland); INFN (Italy); MSIT and NRF (Republic of Korea); MES (Latvia); LMTLT (Lithuania); MOE and UM (Malaysia); BUAP, CINVESTAV, CONACYT, LNS, SEP, and UASLP-FAI (Mexico); MOS (Montenegro); MBIE (New Zealand); PAEC (Pakistan); MES, NSC, and NAWA (Poland); FCT (Portugal); MESTD (Serbia); MICIU/AEI and PCTI (Spain); MOSTR (Sri Lanka); Swiss Funding Agencies (Switzerland); MST (Taipei); MHESI (Thailand); TUBITAK and TENMAK (T\"{u}rkiye); NASU (Ukraine); STFC (United Kingdom); DOE and NSF (USA).

\hyphenation{Rachada-pisek} Individuals have received support from the Marie-Curie program and the European Research Council and Horizon 2020 Grant, contract Nos.\ 675440, 724704, 752730, 758316, 765710, 824093, 101115353, 101002207, 101001205, and COST Action CA16108 (European Union); the Leventis Foundation; the Alfred P.\ Sloan Foundation; the Alexander von Humboldt Foundation; the Science Committee, project no. 22rl-037 (Armenia); the Fonds pour la Formation \`a la Recherche dans l'Industrie et dans l'Agriculture (FRIA) and Fonds voor Wetenschappelijk Onderzoek contract No. 1228724N (Belgium); the Beijing Municipal Science \& Technology Commission, No. Z191100007219010, the Fundamental Research Funds for the Central Universities, the Ministry of Science and Technology of China under Grant No. 2023YFA1605804, the Natural Science Foundation of China under Grant No. 12535004, and USTC Research Funds of the Double First-Class Initiative No.\ YD2030002017 (China); the Ministry of Education, Youth and Sports (MEYS) of the Czech Republic; the Shota Rustaveli National Science Foundation, grant FR-22-985 (Georgia); the Deutsche Forschungsgemeinschaft (DFG), among others, under Germany's Excellence Strategy -- EXC 2121 ``Quantum Universe" -- 390833306, and under project number 400140256 - GRK2497; the Hellenic Foundation for Research and Innovation (HFRI), Project Number 2288 (Greece); the Hungarian Academy of Sciences, the New National Excellence Program - \'UNKP, the NKFIH research grants K 131991, K 133046, K 138136, K 143460, K 143477, K 146913, K 146914, K 147048, 2020-2.2.1-ED-2021-00181, TKP2021-NKTA-64, and 2025-1.1.5-NEMZ\_KI-2025-00004 (Hungary); the Council of Science and Industrial Research, India; ICSC -- National Research Center for High Performance Computing, Big Data and Quantum Computing, FAIR -- Future Artificial Intelligence Research, and CUP I53D23001070006 (Mission 4 Component 1), funded by the NextGenerationEU program (Italy); the Latvian Council of Science; the Ministry of Education and Science, project no. 2022/WK/14, and the National Science Center, contracts Opus 2021/41/B/ST2/01369, 2021/43/B/ST2/01552, 2023/49/B/ST2/03273, and the NAWA contract BPN/PPO/2021/1/00011 (Poland); the Funda\c{c}\~ao para a Ci\^encia e a Tecnologia (Portugal); the National Priorities Research Program by Qatar National Research Fund; MICIU/AEI/10.13039/501100011033, ERDF/EU, ``European Union NextGenerationEU/PRTR", and Programa Severo Ochoa del Principado de Asturias (Spain); the Chulalongkorn Academic into Its 2nd Century Project Advancement Project, the National Science, Research and Innovation Fund program IND\_FF\_68\_369\_2300\_097, and the Program Management Unit for Human Resources \& Institutional Development, Research and Innovation, grant B39G680009 (Thailand); the Eric \& Wendy Schmidt Fund for Strategic Innovation through the CERN Next Generation Triggers project under grant agreement number SIF-2023-004; the Kavli Foundation; the Nvidia Corporation; the SuperMicro Corporation; the Welch Foundation, contract C-1845; and the Weston Havens Foundation (USA).

\end{acknowledgments}

\bibliography{auto_generated}
\cleardoublepage \appendix\section{The CMS Collaboration \label{app:collab}}\begin{sloppypar}\hyphenpenalty=5000\widowpenalty=500\clubpenalty=5000\cmsinstitute{Yerevan Physics Institute, Yerevan, Armenia}
{\tolerance=6000
A.~Hayrapetyan, V.~Makarenko\cmsorcid{0000-0002-8406-8605}, A.~Tumasyan\cmsAuthorMark{1}\cmsorcid{0009-0000-0684-6742}
\par}
\cmsinstitute{Institut f\"{u}r Hochenergiephysik, Vienna, Austria}
{\tolerance=6000
W.~Adam\cmsorcid{0000-0001-9099-4341}, J.W.~Andrejkovic, L.~Benato\cmsorcid{0000-0001-5135-7489}, T.~Bergauer\cmsorcid{0000-0002-5786-0293}, M.~Dragicevic\cmsorcid{0000-0003-1967-6783}, C.~Giordano\cmsorcid{0000-0001-6317-2481}, P.S.~Hussain\cmsorcid{0000-0002-4825-5278}, M.~Jeitler\cmsAuthorMark{2}\cmsorcid{0000-0002-5141-9560}, N.~Krammer\cmsorcid{0000-0002-0548-0985}, A.~Li\cmsorcid{0000-0002-4547-116X}, D.~Liko\cmsorcid{0000-0002-3380-473X}, I.~Mikulec\cmsorcid{0000-0003-0385-2746}, J.~Schieck\cmsAuthorMark{2}\cmsorcid{0000-0002-1058-8093}, R.~Sch\"{o}fbeck\cmsAuthorMark{2}\cmsorcid{0000-0002-2332-8784}, D.~Schwarz\cmsorcid{0000-0002-3821-7331}, M.~Shooshtari\cmsorcid{0009-0004-8882-4887}, M.~Sonawane\cmsorcid{0000-0003-0510-7010}, W.~Waltenberger\cmsorcid{0000-0002-6215-7228}, C.-E.~Wulz\cmsAuthorMark{2}\cmsorcid{0000-0001-9226-5812}
% ...
\par}
\vskip\cmsinstskip
\dag:~Deceased\\
$^{1}$Also at Yerevan State University, Yerevan, Armenia\\
$^{2}$Also at TU Wien, Vienna, Austria\\
% ...
\end{sloppypar}
\end{document}